%% file: HIN-12-011_temp.tex
\pdfoutput=1

\documentclass[11pt,twoside,a4paper,cmspaper,final,collab]{cms-tdr}
\def\svnVersion{229122}\def\cmsCernNo{2013-214}\def\cmsCernDate{\today}\def\cmsMessage{Published in the Journal of High Energy Physics as \href{http://dx.doi.org/10.1007/JHEP02(2014)088}{\doi{10.1007/JHEP02(2014)088.}}}

\begin{document}\cmsNoteHeader{HIN-12-011}

\hyphenation{had-ron-i-za-tion}
\hyphenation{cal-or-i-me-ter}
\hyphenation{de-vices}

\RCS$Revision: 228292 $
\RCS$HeadURL: svn+ssh://svn.cern.ch/reps/tdr2/papers/HIN-12-011/trunk/HIN-12-011.tex $
\RCS$Id: HIN-12-011.tex 228292 2014-02-18 21:39:54Z davidlw $

\newcommand {\ptlow}       {\ensuremath{p_\mathrm{T}^{\text{low}}}}
\newcommand {\ptref}       {\ensuremath{p_\mathrm{T}^{\text{ref}}}}
\newcommand {\PbPb}  {\ensuremath{\mathrm{PbPb}}\xspace}
\newcommand {\roots}    {\ensuremath{\sqrt{s}}}
\newcommand {\ptass}       {\ensuremath{p_\mathrm{T}^{\text{assoc}}}}
\newcommand {\rootsNN}  {\ensuremath{\sqrt{s_{_{NN}}}}}
\newcommand {\pttrg}       {\ensuremath{p_\mathrm{T}^{\text{trig}}}}
\newcommand {\deta}     {\ensuremath{\Delta\eta}}
\newcommand {\dphi}     {\ensuremath{\Delta\phi}}
\newcommand {\npart}    {\ensuremath{\mathrm{N}_\text{Part}}}
\newcommand {\avg}[1]{\ensuremath{\langle\kern-1.0pt\langle#1\rangle\kern-1.0pt\rangle}}
\providecommand{\mubinv} {\ensuremath{\,\mu\mathrm{b}^{-1}}\xspace}
\providecommand{\rd}{\ensuremath{\cmsSymbolFace{d}}}

\cmsNoteHeader{HIN-12-011} 
\title{Studies of azimuthal dihadron correlations in ultra-central \PbPb\ collisions at \rootsNN\ = 2.76\TeV}

\date{\today}

\abstract{
Azimuthal dihadron correlations of charged particles have been measured in \PbPb\
collisions at \rootsNN\ = 2.76\TeV by the CMS collaboration, using data from the 2011 LHC heavy-ion
run. The data set includes a sample of ultra-central (0--0.2\% centrality) \PbPb\ events collected using a
trigger based on total transverse energy in the hadron forward calorimeters
and the total multiplicity of pixel clusters in the silicon pixel tracker. A total
of about 1.8 million ultra-central events were recorded, corresponding to an integrated
luminosity of 120\mubinv. The observed correlations in ultra-central \PbPb\ events
are expected to be particularly sensitive to initial-state fluctuations.
The single-particle anisotropy Fourier harmonics, from $v_2$ to $v_6$, are extracted
as a function of particle transverse momentum. At higher transverse momentum, the
$v_2$ harmonic becomes significantly smaller than the higher-order $v_n$ ($n \geq 3$).
The $\pt$-averaged $v_2$ and $v_3$ are found to be equal within 2\%, while higher-order
$v_n$ decrease as $n$ increases. The breakdown of factorization of dihadron correlations
into single-particle azimuthal anisotropies is observed. This effect is found to be most
prominent in the ultra-central \PbPb\ collisions, where the initial-state fluctuations
play a dominant role. A comparison of the factorization data to hydrodynamic predictions with
event-by-event fluctuating initial conditions is also presented.
}

\hypersetup{%
pdfauthor={CMS Collaboration},%
pdftitle={Studies of azimuthal dihadron correlations in ultra-central PbPb collisions at sqrt(s[NN]) = 2.76 TeV},%
pdfsubject={CMS},%
pdfkeywords={CMS, physics, heavy ions, correlations, harmonic flow}}

\maketitle 

\section{Introduction}
\label{sec_intro}

The azimuthal anisotropy of emitted charged particles is an important feature of
the hot, dense medium produced in heavy-ion collisions. One of the main goals of
studying the azimuthal anisotropies is to understand the
collective properties of the medium and extract its transport coefficients, particularly
the shear viscosity over entropy density ratio, $\eta/s$, using hydrodynamic
models~\cite{Heinz:2013th}. Earlier observations of strong azimuthal anisotropies
in collisions of gold nuclei at nucleon-nucleon center-of-mass energies (\rootsNN) up
to 200\GeV at the Relativistic Heavy-Ion Collider (RHIC)
indicated that a strongly coupled quark-gluon plasma is produced, which behaves
as a nearly perfect liquid with a close-to-zero $\eta/s$
value~\cite{PHENIX,STAR,PHOBOS,BRAHMS,Shuryak:2004cy,Gyulassy:2004zy}.
The azimuthal anisotropies have also been extensively measured
at the Large Hadron Collider (LHC) over a wide kinematic range in
\PbPb\ collisions at \rootsNN\ =
2.76\TeV~\cite{Chatrchyan:2012ta,Chatrchyan:2012wg,Chatrchyan:2012xq,Chatrchyan:2012vqa,Aamodt:2010pa,ALICE:2011ab,Aamodt:2011by,ATLAS:2011ah,ATLAS:2012at,Aad:2013xma}.

In a non-central heavy-ion collision, the overlap region of the two colliding
nuclei has a lenticular shape, and the interacting nucleons in this region are
known as ``participants.'' The ``participant plane'' is defined by the beam
direction and the short axis of the participating nucleon distribution.
Because of fluctuations that arise from the finite number of nucleons,
the impact parameter vector typically does not coincide with
the short axis of this lenticular region. Strong
rescattering of the partons in the initial state may lead to local thermal
equilibrium and the build-up of anisotropic pressure gradients, which drive a
collective anisotropic expansion. The expansion is fastest along the largest
pressure gradient, i.e., along the short axis of the lenticular region.
Therefore, the eccentricity of initial-state collision geometry
results in an anisotropic azimuthal distribution of the final-state hadrons.
In general, the anisotropy can be characterized by the Fourier harmonic coefficient
($v_{n}$) in the azimuthal angle ($\phi$) distribution of the hadron yield,
$\rd{}N/\rd\phi \propto 1+2 \sum_{n}^{} v_{n}\cos[n(\phi-\Psi_\mathrm{n})]$,
where $\Psi_\mathrm{n}$ is the event-by-event azimuthal angle of the participant
plane. As the participant plane is not a measurable quantity experimentally,
it is often approximated by the ``event plane'', defined as the direction of
maximum final-state particle density. The second-order Fourier component ($v_2$)
is known as the ``elliptic flow'', and its event plane angle $\Psi_{2}$
approximately corresponds to the short axis direction of the lenticular region.
Due to event-by-event fluctuations, higher-order deformations or eccentricities
of the initial geometry can also be induced, which lead to higher-order Fourier
harmonics ($v_n$, $n \ge 3$) in the final state with respect to their corresponding event plane angles,
$\Psi_\mathrm{n}$~\cite{PhysRevLett.104.142301,PHOBOSeccPART,Bhalerao:2006tp,Voloshin:2007pc,Ollitrault:2009ie,Alver:2010gr,Qiu:2011iv}.
For a given initial-state eccentricity, the finite $\eta/s$ value of the system
tends to reduce the azimuthal anisotropy observed for final-state particles. The higher-order Fourier
harmonics are expected to be particularly sensitive to the shear viscosity of the
expanding medium.

Precise extraction of $\eta/s$ from the anisotropy data is crucial
for investigating the transport properties of the hot and dense medium created in
heavy-ion collisions in detail~\cite{Heinz:2013th}. This effort is, however, complicated by
large uncertainties in our understanding of the initial-state conditions of
heavy-ion collisions, especially in terms of event-by-event fluctuations.
Different initial-state models predict different values of eccentricity and
its fluctuations, leading to large uncertainties on the extracted
$\eta/s$ values. In order to better constrain the initial-state condition,
it was suggested~\cite{Luzum:2012wu} that in ultra-central
heavy-ion collisions (e.g., top 1\% most central collisions),
the initial collision geometry is predominantly generated
by fluctuations such that various orders of eccentricities predicted by different
models tend to converge. Here, collision centrality is defined as the fraction of
the total inelastic \PbPb\ cross section, with 0\% denoting the most central collisions.
Therefore, studies of azimuthal anisotropy in ultra-central
heavy-ion collisions can help to reduce the systematic uncertainties of initial-state
modeling in extracting the $\eta/s$ value of the system, although
quantitative comparison to theoretical calculations is beyond the scope of this paper.

Furthermore, since the event plane angle, $\Psi_\mathrm{n}$, is determined by the
final-state particles, selecting particles from different ranges of transverse
momentum (\pt) may lead to different estimates of event plane angles.
Also due to the effect of initial-state fluctuations, it was recently predicted
by hydrodynamic models that a $\pt$-dependence of the event plane angle will be induced,
which could be one of the sources responsible for the
breakdown of factorization in extracting $v_n$ harmonics from
dihadron correlations~\cite{Gardim:2012im,Heinz}. As mentioned already,
the ultra-central heavy-ion events are dominated by the initial-state
eccentricity fluctuations. Thus, they provide an ideal testing ground for the
effect of a $\pt$-dependent event plane angle.

This paper presents the measurement of azimuthal anisotropy harmonics, from $v_2$ to $v_6$,
extracted using long-range (large $\abs{\deta}$) dihadron correlations as a function of
\pt from 0.3 to 8.0\GeVc in the top 0.2\% most central \PbPb\ collisions at a
center-of-mass energy per nucleon pair (\rootsNN) of 2.76\TeV.
Here, $\deta$ is the difference in pseudorapidity $\eta$ =$-\ln[\tan(\theta/2)]$
between the two particles, where the polar angle $\theta$ is defined relative
to the beam axis. The $\pt$-averaged $v_n$ values for $0.3<\pt<3.0$\GeVc are also
derived up to $n=7$. Factorization of the Fourier coefficients from dihadron correlations
into a product of single-particle azimuthal anisotropies is investigated. This study
of factorization is quantitatively compared to hydrodynamic predictions with different
models of initial-state fluctuations and $\eta/s$ values for two centrality classes.

\section{Experimental Setup}
\label{sec:Data_selection}

The data used in this analysis correspond to an integrated luminosity of
120\mubinv and were recorded with the CMS detector during the 2011 \PbPb\
LHC running period at \rootsNN\ =  2.76\TeV. A detailed description of the CMS
detector can be found in Ref.~\cite{CMS:2008zzk}.
The CMS uses a right-handed coordinate system, with the origin at the nominal
interaction point, the $x$ axis pointing to the centre of the LHC, the $y$
axis pointing up (perpendicular to the LHC plane), and the $z$ axis
along the anticlockwise-beam direction. The polar angle $\theta$ is
measured from the positive $z$ axis and the azimuthal angle ($\phi$) is measured
in the $x$-$y$ plane. The central feature of the apparatus is
a superconducting solenoid of 6\unit{m} internal diameter, providing a
magnetic field of 3.8\unit{T}. Within the field volume are the silicon
pixel and strip trackers, the crystal electromagnetic calorimeter,
and the brass/scintillator hadron calorimeter.
In \PbPb\ collisions, trajectories of charged particles with $\pt > 0.2$\GeVc are
reconstructed in the tracker covering the pseudorapidity
region $\abs{\eta}<2.5$, with a track momentum resolution of about 1\% at $\pt = 100\GeV/c$.
In addition, CMS has extensive forward calorimetry, in
particular two steel/quartz-fiber Cherenkov hadron forward (HF)
calorimeters, which cover the pseudorapidity range $2.9 < \abs{\eta} < 5.2$.
The HF calorimeters are segmented into towers, each of which is a two-dimensional
cell with a granularity of 0.5 units in $\eta$ and 0.349 rad in $\phi$.
The zero-degree calorimeters (ZDC) are tungsten/quartz Cherenkov
calorimeters located at $\pm$140\mm from the interaction point~\cite{Grachov:2008qg}.
They are designed to measure the energy of photons and spectator neutrons
emitted from heavy ion collisions.
Each ZDC calorimeter has electromagnetic and hadronic sections with an
active area of $\pm$40\mm in x and $\pm$50\mm in y. When the LHC beam
crossing angle is 0 degree, this corresponds to an $\eta$ acceptance that
starts at $\eta=8.3$ and is 100\% by $\eta=8.9$ for \rootsNN\ = 2.76\TeV.
For one neutron, the ZDCs have an energy resolution of 20\%. Since each
neutron interacts independently, the resolution improves as the square root
of the number of neutrons.

\section{Selections of Events and Tracks}
\label{subsec_trigger_minbias}

Minimum bias \PbPb\ events were triggered by coincident signals from both
ends of the detector in either the beam scintillator counters (BSC) at
$3.23 < \abs{\eta} < 4.65$ or in the HF calorimeters. Events due to noise,
cosmic rays, out-of-time triggers, and beam backgrounds were suppressed
by requiring a coincidence of the minimum bias trigger with bunches colliding
in the interaction region. The trigger has an efficiency of $(97 \pm 3)$\%
for hadronic inelastic \PbPb\ collisions. In total, about 2\% of all
minimum bias \PbPb\ events were recorded.

\begin{figure}[thb]
  \begin{center}
    \includegraphics[width=0.7\linewidth]{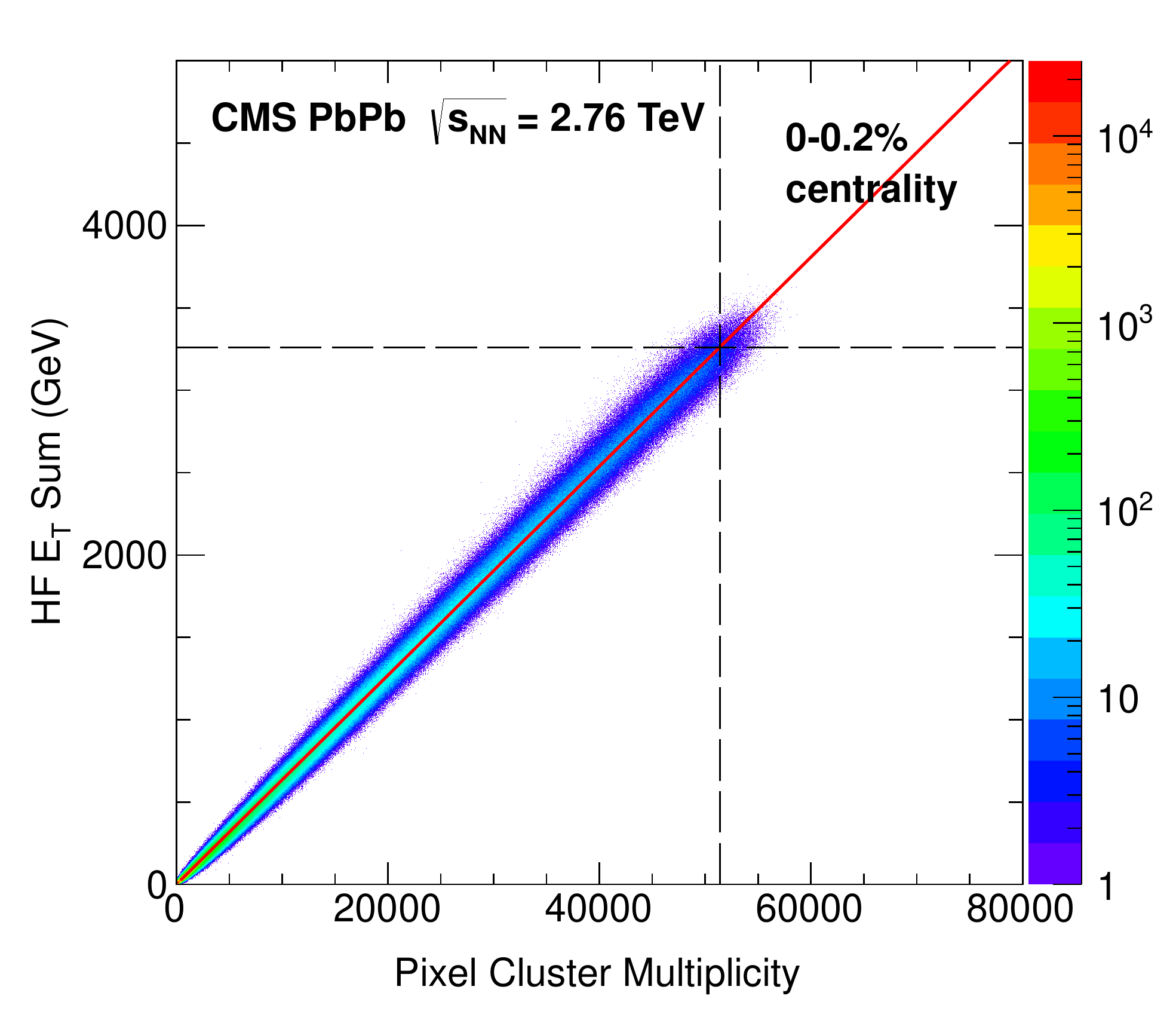}
    \caption{HF \ET\ sum vs. pixel cluster multiplicity for minimum bias triggered
    \PbPb\ collisions at \rootsNN\ = 2.76\TeV. The region in the upper right corner
    encompassed by the dashed lines depicts the 0--0.2\% selected centrality range.
}
    \label{fig:HFVsPixel_mb_nopurej}
  \end{center}
\end{figure}

To maximize the event sample for very central \PbPb\ collisions, a dedicated online
trigger on the 0--0.2\% ultra-central events was implemented by simultaneously requiring the HF
transverse energy (\ET) sum to be greater than 3260\GeV and the pixel cluster multiplicity to be greater than
51400 (which approximately corresponds to 9500 charged particles
over 5 units of pseudorapidity). The selected events correspond
to the 0.2\% most central collisions of the total \PbPb\
inelastic cross section. The correlation between the HF \ET\ sum and pixel
cluster multiplicity for minimum bias \PbPb\ collisions at \rootsNN\ =
2.76\TeV is shown in Fig.~\ref{fig:HFVsPixel_mb_nopurej}. The dashed lines
indicate the selections used for the 0--0.2\% centrality range. This fractional
cross section is determined relative to the standard 0--2.5\% centrality selection
in \PbPb\ collisions at CMS by selecting on the total energy deposited in the
HF calorimeters~\cite{Chatrchyan:2012ta}.
The inefficiencies of the minimum bias trigger and event
selection for very peripheral events are properly accounted. In a similar way,
the 0--0.02\% centrality range is also determined by requiring the HF \ET\
sum greater than 3393\GeV and pixel cluster multiplicity
greater than 53450 (a subset of 0--0.2\% ultra-central events).
With this trigger, the ultra-central \PbPb\ event sample is
enhanced by a factor of about 40 compared to the minimum bias sample.
For purposes of systematic comparisons, other \PbPb\ centrality ranges,
corresponding to 40--50\%, 0--10\%, 2.5--5.0\%, 0--2.5\%
and 0--1\%, are studied based on the HF \ET\ sum selection using the minimum bias sample.
As a cross-check, the 0--1\% centrality range is also studied using
combined HF \ET\ sum and pixel cluster multiplicity, similar to the centrality
selection of 0--0.2\% ultra-central events.

Centrality selections of ultra-central events are investigated in
Monte Carlo (MC) simulations using the \textsc{ampt}~\cite{Lin:2004en} heavy-ion
event generator, which provides a realistic modeling of the initial-state fluctuations of
participating nucleons. The generated particles are propagated through the
full \textsc{geant4}~\cite{GEANT4} simulation of the CMS detector. The equivalent centrality requirements
on the HF \ET\ sum and pixel cluster multiplicity are applied in order
to evaluate the selected ranges of impact parameter and number of participating
nucleons, \npart, for various centrality ranges. A summary of the
mean and RMS values of \npart\ distributions for selected events of each very central
\PbPb\ centrality range can be found in Table~\ref{table:npart_ampt}.
As one can see, there is only a moderate increase of average \npart\
value for events that are more central than 0--1\% centrality, although the RMS value still decreases
significantly for more central selections.

\begin{table}[htb]
\centering
\caption{\label{table:npart_ampt} The mean and RMS of \npart\ distributions
for selected events in each centrality bin in \textsc{ampt} simulations.}
\vspace{2mm}
\begin{tabular}{c|c|c}
\hline
\hline
Centrality & $\langle \npart\ \rangle$ & RMS \\
\hline
0--0.02\% & 406.2 & 3.6 \\
0--0.2\% & 404.0 & 6.9 \\
0--1.0\% & 401.1 & 8.3 \\
0--2.5\% & 395.8 & 11.3 \\
2.5--5.0\% & 381.3 & 19.5 \\
\hline
\hline
\end{tabular}
\end{table}

Standard offline event selections~\cite{Chatrchyan:2012ta} are also applied by requiring energy deposits
in at least three towers in each of the HF calorimeters, with
at least 3\GeV of energy in each tower, and the presence of a reconstructed
primary vertex containing at least two tracks. The reconstructed
primary vertex is required to be located within $\pm$15~cm of the average
interaction region along the beam axis and within a radius of 0.02~cm
in the transverse plane. These criteria further reduce the
background from single-beam interactions (e.g., beam-gas and beam-halo),
cosmic muons, and ultra peripheral
collisions that lead to the electromagnetic breakup of one or
both Pb nuclei~\cite{Djuvsland:2010qs}. These
criteria are most relevant for selecting very peripheral \PbPb\ events
but have little effect ($<0.01$\%) on the events studied in this paper.

During the 2011 \PbPb\ run, there was a probability of about $10^{-3}$ to
have two collisions recorded in a single beam crossing (pileup events).
This probability is even higher for ultra-central triggered events,
which sample the tails of the HF \ET\ sum and pixel cluster multiplicity
distributions. If a large HF \ET\ sum or pixel cluster multiplicity
event is due to two mid-central collisions instead of a single ultra-central
collision, more spectator neutrons will be
released, resulting in a large signal in the ZDC. To select cleaner
single-collision \PbPb\ events, the correlation of energy sum signals
between ZDC and HF detectors is studied. Events with large signals
in both ZDC and HF are identified as pileup events (about 0.1\% of all events), and thus rejected.

The reconstruction of the primary event vertex and the trajectories of charged
particles in \PbPb\ collisions is based on signals in the silicon pixel and strip detectors
and described in detail in Ref.~\cite{Chatrchyan:2012ta}.
From studies based on \PbPb\ events simulated
using \textsc{hydjet}~\cite{Lokhtin:2005px} (version 1.8), the combined geometrical
acceptance and reconstruction efficiency of the primary tracks is about 70\%
at $\pt\sim1\GeVc$ and $\abs{\eta} < 1.0$ for the 0--0.2\% central \PbPb\ events
but drops to about 50\% for $\pt \sim 0.3$\GeVc.
The fraction of misidentified tracks is kept at the level of $<5$\% over
most of the \pt ($\pt>0.5$\GeVc) and $\eta$ ($\abs{\eta}<1.6$) ranges.
It increases up to about 20\% for very low \pt ($\pt<0.5\GeVc$) particles in
the forward ($\abs{\eta} \approx 2)$ region.

\section{Analysis procedure}
\label{sec:analysis}

Following the same procedure of dihadron correlation analysis as in
Refs.~\cite{Khachatryan:2010gv,Chatrchyan:2011eka,Chatrchyan:2012wg,CMS:2012qk,Chatrchyan:2013nka},
the signal and background distributions of particle pairs are first constructed.
Any charged particle associated with the primary vertex and in the range $\abs{\eta} < 2.4$
can be used as a ``trigger'' particle. A variety of bins of trigger particle transverse momentum,
denoted by \pttrg, are considered. In a single event, there can be more than one trigger particle
and their total multiplicity is denoted
by $N_\text{trig}$. Within each event, every trigger particle is then paired with
all of the remaining particles (again within $\abs{\eta} < 2.4$). Just as for the trigger
particles, these associated particles are also binned in transverse momentum (\ptass).

The signal distribution, $S(\Delta\eta,\Delta\phi)$, is the per-trigger-particle
yield of pairs found in the same event,
\begin{linenomath}
\begin{equation}
\label{eq:signal}
S(\Delta\eta,\Delta\phi) = \frac{1}{N_\text{trig}}\frac{\rd^{2}N^\text{same}}{\rd\Delta\eta\, \rd\Delta\phi},
\end{equation}
\end{linenomath}
where $N^\text{same}$ is the number of such pairs within a ($\Delta\eta$,$\Delta\phi$) bin, and
$\dphi$ and $\deta$ are the differences in azimuthal angle $\phi$ and pseudorapidity $\eta$
between the two particles. The background distribution, $B(\Delta\eta,\Delta\phi)$, is found using a mixed-event
technique, wherein trigger particles from one event are combined (mixed) with all of
the associated particles from a different event. In the analysis, associated particles
from 10 randomly chosen events with a small $z_\text{vtx}$ range ($\pm$0.5\cm) near the $z_\text{vtx}$
of the event with trigger particles are used. The result is given by

\begin{linenomath}
\begin{equation}
\label{eq:background}
B(\Delta\eta,\Delta\phi) = \frac{1}{N_\text{trig}}\frac{\rd^{2}N^\text{mix}}{\rd\Delta\eta\, \rd\Delta\phi},
\end{equation}
\end{linenomath}

\noindent where $N^\text{mix}$ denotes the number of
mixed-event pairs. This background distribution represents the
expected correlation function assuming independent particle emission,
but taking into account effects of the finite acceptance.

The two-dimensional (2D) differential yield of associated particles
per trigger particle is given by
\begin{linenomath}
\begin{equation}
\label{2pcorr_incl}
\frac{1}{N_\text{trig}}\frac{\rd^{2}N^\text{pair}}{\rd\Delta\eta\, \rd\Delta\phi}
= B(0,0)\times\frac{S(\Delta\eta,\Delta\phi)}{B(\Delta\eta,\Delta\phi)},
\end{equation}
\end{linenomath}
\noindent where $N^\text{pair}$ is the total number of hadron pairs.
The value of the background distribution at $\Delta\eta=0$ and $\Delta\phi=0$,
$B(0,0)$, represents the mixed-event associated yield for
both particles of the pair going in approximately the same direction and
thus having full pair acceptance (with a bin width of 0.3 in $\Delta\eta$ and
$\pi/16$ in $\Delta\phi$). Therefore, the ratio $B(0,0)/B(\Delta\eta,\Delta\phi)$
accounts for the pair-acceptance effects. The correlation function described in
Eq.~(\ref{2pcorr_incl}) is calculated in 0.5~cm wide bins of the $z_{\rm vtx}$
along the beam direction and then averaged over the range $\abs{z_\text{vtx}} < 15\cm$.

To extract the azimuthal anisotropy harmonics, $v_n$, the one-dimensional
(1D) azimuthal dihadron correlation function as a function of \dphi, averaged
over $\abs{\deta}>2$ (to avoid the short-range correlations from jets and resonance
decays), can be decomposed into a Fourier series given by
\begin{linenomath}
\begin{equation}
\label{fourier}
\frac{1}{N_\text{trig}}\frac{\rd N^\text{pair}}{\rd\Delta\phi} = \frac{N_\text{assoc}}{2\pi} \left\{ 1+\sum\limits_{n=1}^{\infty} 2V_{n\Delta} \cos (n\Delta\phi)\right\}.
\end{equation}
\end{linenomath}
\noindent Here, $V_{n\Delta}$ are the Fourier coefficients from dihadron correlations, and
$N_\text{assoc}$ represents the total number of hadron pairs per trigger
particle for a given $|\Delta\eta|$ range and $(\pttrg, \ptass)$ bin.

In Refs.~\cite{Chatrchyan:2011eka,Chatrchyan:2012wg,CMS:2012qk,Chatrchyan:2013nka},
a fit to the azimuthal correlation function by a Fourier series was used
to extract the $V_{n\Delta}$ coefficients. In this paper, a slightly different approach
is applied. The $V_{n\Delta}$ values are directly calculated as the average
value of $\cos(n\Delta\phi)$ of all particle pairs for $ \abs{\deta} > 2$
(to avoid the short-range correlations from jets and resonance decays):
\begin{linenomath}
\begin{equation}
\label{average}
V_{n\Delta}=\avg{\cos(n\Delta\phi)}_{S}-\avg{\cos(n\Delta\phi)}_{B}.
\end{equation}
\end{linenomath}
Here, $\avg{\ }$ denotes averaging over all
particles in each event and over all the events. The subscripts $S$ and $B$
correspond to the average over signal and background pairs. With an ideal detector,
$\avg{\cos(n\Delta\phi)}_{S}$ equals to $V_{n\Delta}$ by definition.
The $\avg{\cos(n\Delta\phi)}_{B}$ term is subtracted in order to remove
the effects of detector non-uniformity. The advantage
of the present approach is that the extracted Fourier harmonics will not be affected
by the finite bin widths of the histogram in $\Delta\eta$ and $\Delta\phi$.
This is particularly important for very-high-order harmonics ($V_{n\Delta}$
is extracted up to $n=7$ in this analysis) that are sensitive to the finer
variations of the correlation functions.

It was thought~\cite{Chatrchyan:2012wg,ATLAS:2012at,Aamodt:2011by} that,
for correlations purely driven by the hydrodynamic flow, $V_{n\Delta}$ can
be factorized into a product of single-particle Fourier harmonics, $v_{n}(\pttrg)$,
for trigger particles and $v_{n}(\ptass)$, for associated particles:
\begin{linenomath}
\begin{equation}
\label{eq:factorization}
V_{n\Delta}=v_{n}(\pttrg) \times v_{n}(\ptass).
\end{equation}
\end{linenomath}
\noindent The single-particle azimuthal anisotropy harmonics can then be
extracted as a function of \pt as follows:
\begin{linenomath}
\begin{equation}
\label{eq:vn_factorization}
v_{n}(\pt)=\frac{V_{n\Delta}(\pt,\ptref)}{\sqrt{V_{n\Delta}(\ptref,\ptref)}} ,
\end{equation}
\end{linenomath}
where a fixed \ptref\ range is chosen for the ``reference particles''.
However, as pointed out in Refs.~\cite{Gardim:2012im,Heinz}, due to
fluctuating initial-state geometry, the factorization of $V_{n\Delta}$ could
also break down for flow-only correlations. Direct tests of the factorization
relation for $V_{n\Delta}$ in Eq.~(\ref{eq:factorization}) are carried out in this
paper, as will be discussed in Section~\ref{sec:factorization}. These tests may provide
new insights into the initial-state density fluctuations of the expanding hot medium.

When calculating $\avg{\cos(n\Delta\phi)}$, each pair
is weighted by the product of correction factors for the two particles. These
factors are the inverse of an efficiency that is a function of each particle's
pseudorapidity and transverse momentum,
\begin{linenomath}
\begin{equation}
\varepsilon_\text{trk}(\eta,\pt) = \frac{A(\eta,\pt) E(\eta,\pt)}{1-F(\eta,\pt)},
\end{equation}
\end{linenomath}
where $A(\eta,\pt)$ is the geometrical acceptance, $E(\eta,\pt)$
is the reconstruction efficiency, and $F(\eta,\pt)$ is the fraction of
misidentified tracks. The effect of this weighting factor only changes
the overall scale of dihadron correlation functions, and has almost no effect
on $\avg{\cos(n\Delta\phi)}$. However, the misidentified
tracks may have different $v_n$ values from those of correctly reconstructed tracks.
Therefore, the effects of misidentified tracks are investigated and corrected using the same procedure
as done in Ref.~\cite{Chatrchyan:2012ta}. The $v_n$ values for the true charged
tracks ($v_n^\text{true}$) can be expressed as a combination of $v_n$ for all the
observed tracks ($v_n^\text{obs}$) and for misidentified tracks ($v_n^\text{mis}$):
\begin{linenomath}
\begin{equation}
v_{n}^\text{true}(\pt) = \frac{v_{n}^\text{obs}(\pt)-F(\pt) \times v_{n}^\text{mis}(\pt)}{1-F(\pt)}.
\end{equation}
\end{linenomath}
An empirical correction for the misidentified track $v_n$ based
on the simulation studies is found to be independent of track selections or the
fraction of misidentified tracks. The correction is given by $v_n^\text{mis} = f \times \left\langle v_n\right\rangle$,
where $\left\langle v_n\right\rangle$ is the yield-weighted average over the \pt range from
0.3 to 3.0\GeVc, folding in the efficiency-corrected spectra. The estimated values
of the correction factor, $f$, as well as its uncertainty, are summarized in Table~\ref{tab:scale-table}
for different $v_n$.

\begin{table*}[ht]
\topcaption{\label{tab:scale-table} The factor, $f$, for estimating the $v_n$ values
of misidentified tracks, as well as its uncertainty, for various orders of Fourier harmonics.}

\begin{center}
\begin{tabular}{lc}
\hline
\hline
 $n$      & $f$\\
\hline
 2        & $1.3 \pm 0.1$ \\
 3        & $1.0 \pm 0.4$ \\
 4        & $0.8 \pm 0.6$ \\
 5        & $0.8 \pm 0.6$ \\
 $>$6     & $0.8 \pm 0.6$ \\
\hline
\hline
\end{tabular}
\end{center}
\end{table*}

The systematic uncertainties due to misidentified tracks, which are most important
at low \pt where the misidentified track rate is high, are reflected
in the uncertainty of the $f$ factor in Table~\ref{tab:scale-table}. At low \pt,
the systematic uncertainty from this source is 1.4\% for $v_2$ and 5--8\% for $v_3$ to $v_6$.
By varying the $z$-coordinate of vertex binning in the mixed-event background, the results
of the $v_{n}$ values vary by at most 2--8\% for $v_2$ to $v_6$, respectively. Systematic uncertainties
due to the tracking efficiency correction are estimated to be about 0.5\%. By varying
the requirements on the ZDC sum energy used for pileup rejection, the results are stable
within less than 1\%. The various
sources of systematic uncertainties are added in quadrature to obtain the
final uncertainties shown as the shaded color bands for results in Section~\ref{sec:results}.

\section{Results}
\label{sec:results}

\subsection{Single-particle azimuthal anisotropy, \texorpdfstring{$v_n$}{v[n]}}
\label{sec:vn}

\begin{figure}[thb]
  \begin{center}
    \includegraphics[width=0.48\linewidth]{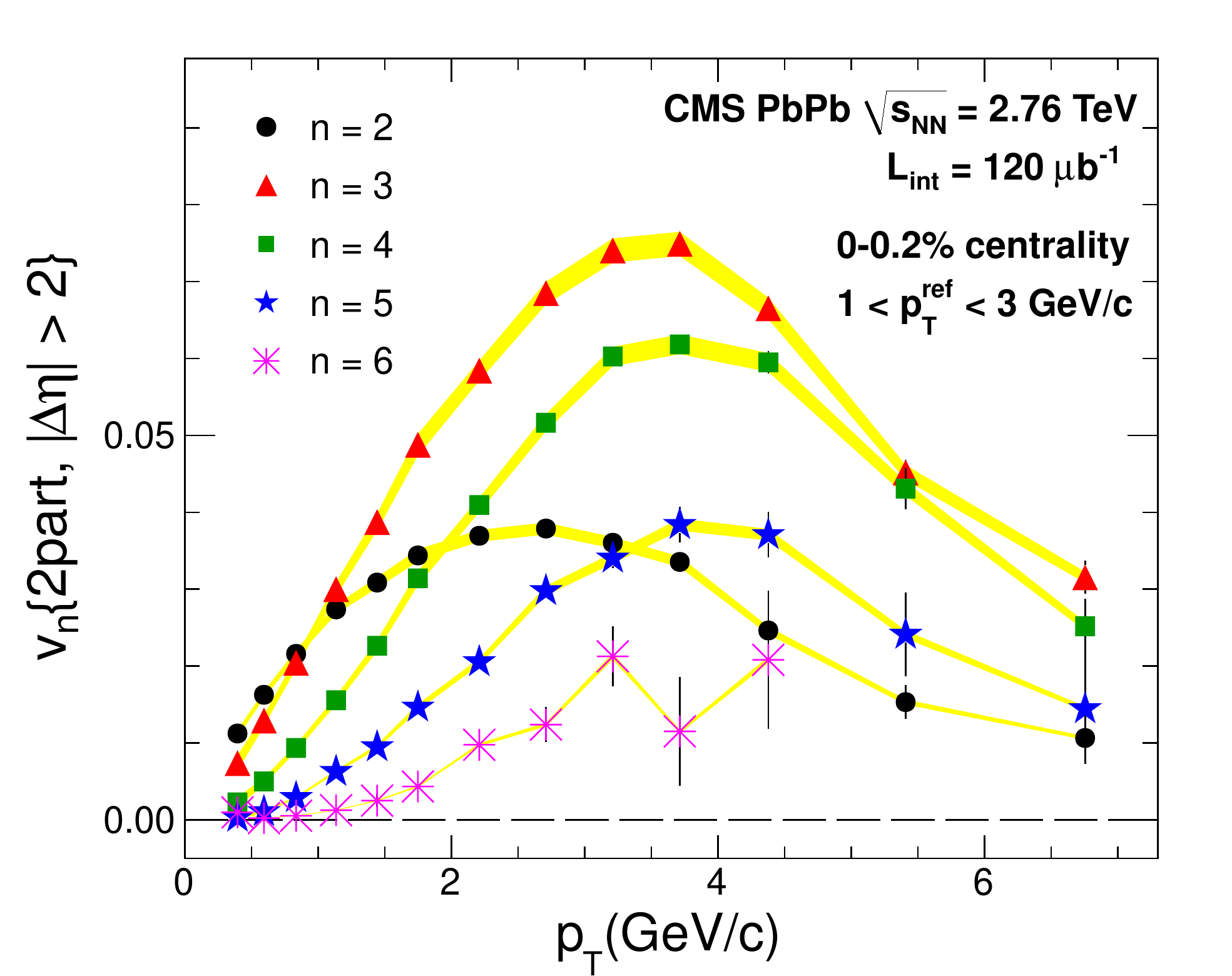}
    \includegraphics[width=0.48\linewidth]{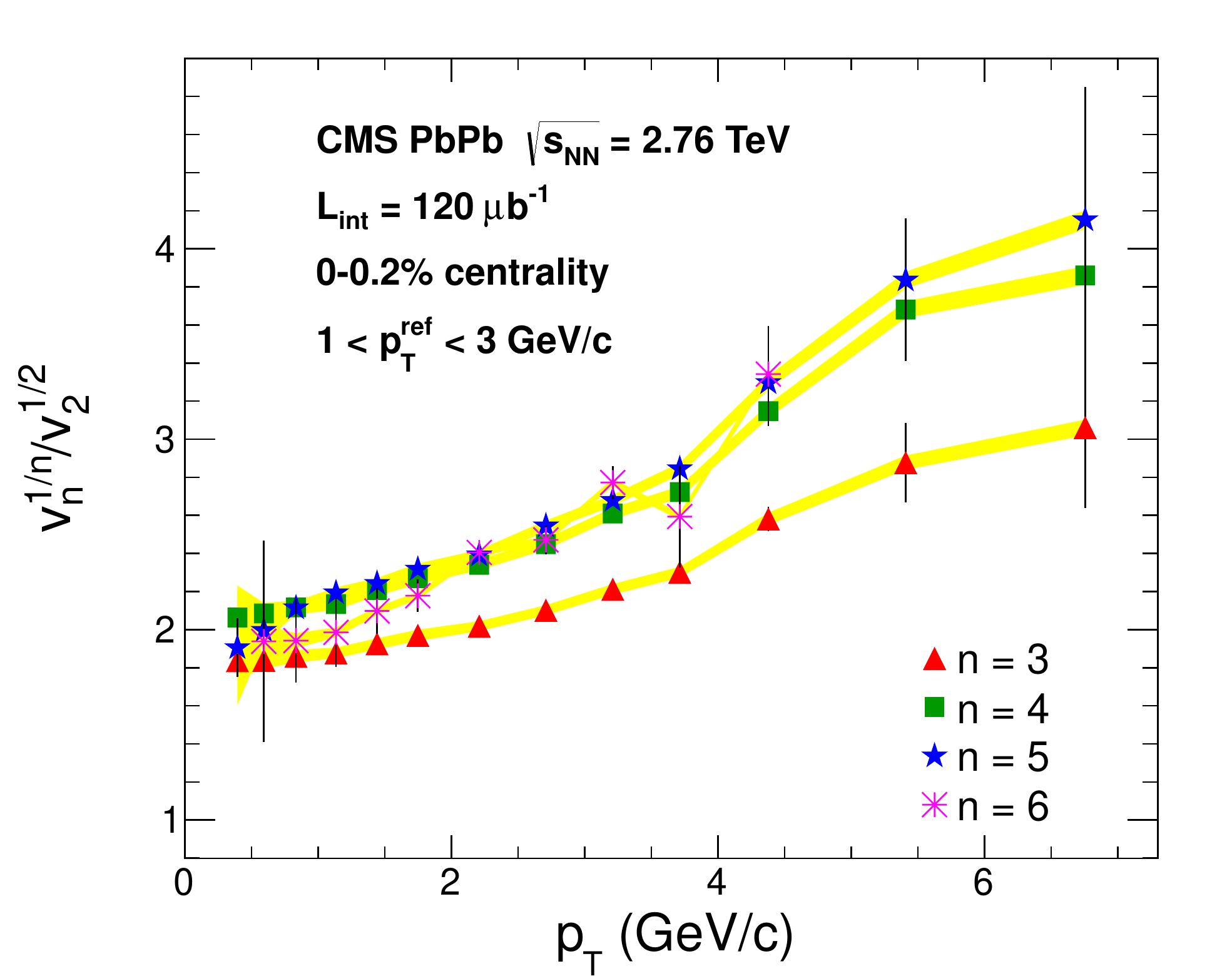}
    \caption{ Left: the $v_2$ to $v_6$ values as a function of \pt in
    0--0.2\% central \PbPb\ collisions at \rootsNN\ = 2.76\TeV. Right:
    the $v^{1/n}_{n}/v^{1/2}_{2}$ ratios as a function of \pt. Error bars
    denote the statistical uncertainties, while the shaded color bands correspond to
    the systematic uncertainties.}
    \label{fig:compare_vnpt_ucc}
  \end{center}
\end{figure}

\begin{figure}[thb]
  \begin{center}
    \includegraphics[width=\linewidth]{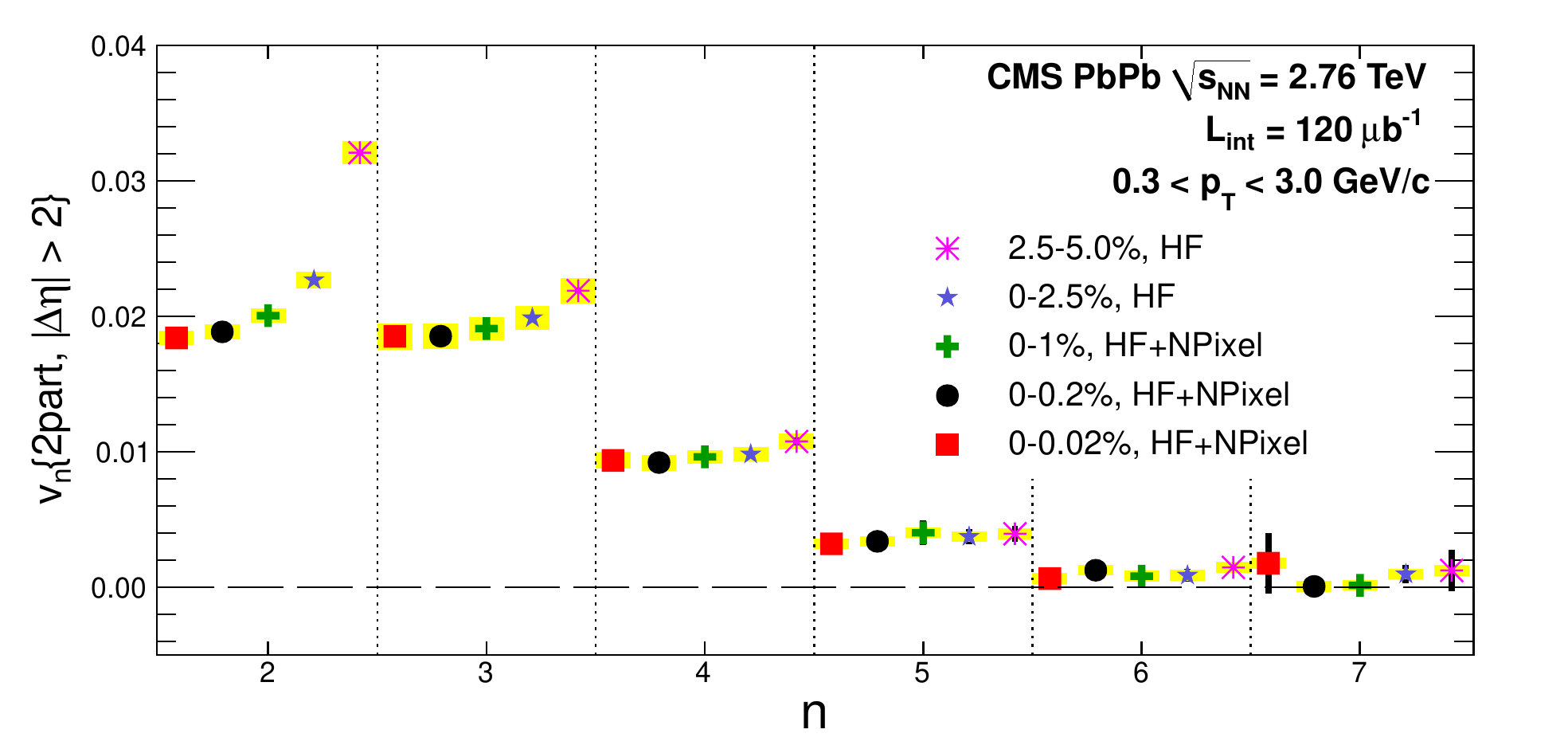}
    \caption{Comparison of $\pt$-averaged (0.3--3.0\GeVc) $v_n$ as a function of $n$
    in five centrality ranges (2.5--5.0\%, 0--2.5\%, 0--1\%, 0--0.2\% and 0--0.02\%)
    for \PbPb\ collisions at \rootsNN\ = 2.76\TeV. The \ptref\ of 1--3\GeVc is used.
    Error bars denote the statistical uncertainties, while the shaded color boxes
    correspond to the systematic uncertainties.}
    \label{fig:compare_vnn_centrality}
  \end{center}
\end{figure}

Results of azimuthal anisotropy harmonics, from $v_2$ to $v_6$, as a function of \pt
in 0--0.2\% central \PbPb\ collisions at \rootsNN\ = 2.76\TeV, are shown in
Fig.~\ref{fig:compare_vnpt_ucc} (left). The $v_n$ values are extracted from
long-range ($\abs{\deta}>2$) dihadron correlations using Eq.~(\ref{average}), and by
assuming factorization in Eq.~(\ref{eq:vn_factorization}). The \ptref\ range is
chosen to be 1--3\GeVc. The error bars correspond to statistical uncertainties,
while the shaded color bands indicate the systematic uncertainties.  As the collisions
are extremely central, the eccentricities, $\epsilon_\mathrm{n}$, are mostly driven
by event-by-event participant fluctuations and are of similar sizes within a few
\% for all orders. Consequently, the magnitudes of $v_2$ and $v_3$ are observed to
be comparable (within 2\% averaged over \pt as will be shown in Fig.~\ref{fig:compare_vnn_centrality}),
which is not the case for non-central collisions. Different $v_n$
harmonics have very different dependencies on \pt. At low \pt ($\pt < 1$\GeVc), the $v_2$
harmonic has the biggest magnitude compared to other higher-order harmonics.
It becomes smaller than $v_3$ at $\pt\approx 1\GeVc$, and even smaller than $v_5$
for $\pt>3$\GeVc. This intriguing \pt dependence can be compared
quantitatively to hydrodynamics calculations with fluctuating initial conditions,
and it provides important constraints on theoretical models. For a given value
of \pt, the magnitude of $v_n$ for $n \geq 3$ decreases monotonically with $n$, as will
be shown later.

\begin{figure}[thb]
  \begin{center}
    \includegraphics[width=0.8\linewidth]{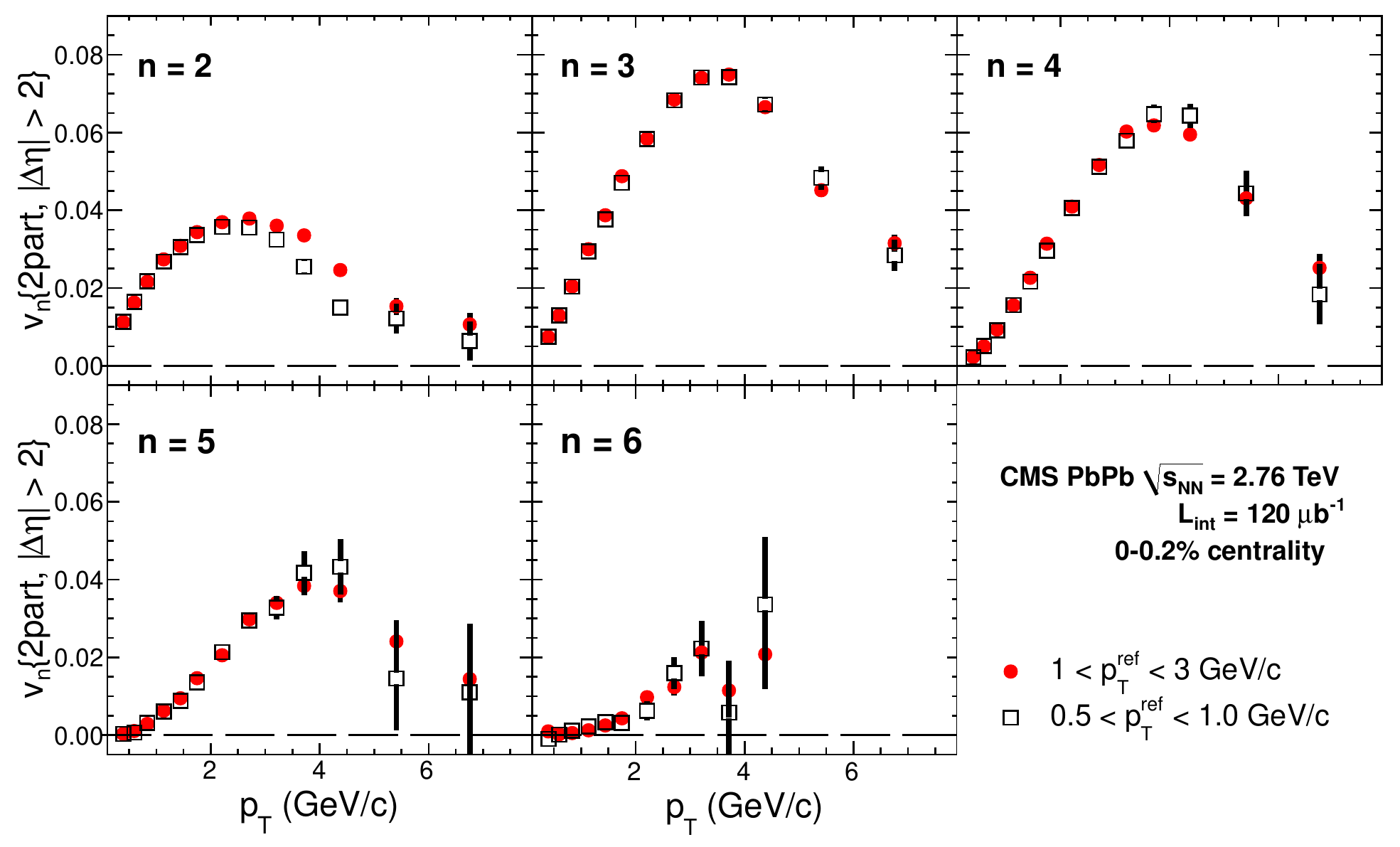}
    \caption{Comparison of $v_{n}(\pt)$ values derived from two different \ptref\
    ranges: 0.5--1.0\GeVc (open square markers) and 1--3\GeVc (solid circles),
    in 0--0.2\% central \PbPb\ collisions at
    \rootsNN\ = 2.76\TeV. Error bars denote the statistical uncertainties.}
    \label{fig:compare_vnpt_lowptass}
  \end{center}
\end{figure}

If a system created in an ultra-relativistic heavy-ion collision behaves
according to ideal hydrodynamics, the Fourier harmonics, $v_{n}$, are expected
to follow a \pt dependence that has a power-law, $\pt^{n}$, functional form
in the low-\pt region~\cite{Alver:2010dn, Borghini:2005kd}.
Hence, the scaling ratio, $v^{1/n}_{n}/v^{1/2}_{2}$, will be largely independent
of \pt, as was seen by the ATLAS collaboration for not very central events~\cite{ATLAS:2012at}.
In Fig.~\ref{fig:compare_vnpt_ucc} (right), the $v^{1/n}_{n}/v^{1/2}_{2}$ ratios
are shown as a function of \pt for $n=3$--$6$ obtained in 0--0.2\% ultra-central
\PbPb\ collisions at \rootsNN\ = 2.76\TeV. The obtained ratio shows an increase
as a function of \pt. This trend is consistent to what was observed by the ATLAS
collaboration for very central events (e.g., 0--1\% centrality)~\cite{ATLAS:2012at}.

Other choices of \ptref\ ranges are also studied in order to examine the assumption
of factorization made for extracting $v_n$. As an example,
Fig.~\ref{fig:compare_vnpt_lowptass} shows the comparison of $v_n$ as a
function of \pt for $1<\ptref<3$\GeVc and $0.5<\ptref<1.0$\GeVc. The $v_n$ values
extracted with two choices of \ptref\ ranges are consistent within statistical
uncertainties for $n>2$ over the entire \pt range. However, a significant discrepancy
is observed for $v_2$ at higher \pt, e.g., up to about 40\% for $\pt \sim 4$\GeVc,
while the low \pt region shows a good agreement between the two \ptref\ ranges.
A detailed study of factorization breakdown for Eq.~(\ref{eq:factorization})
as well as its physical implication is presented in Section~\ref{sec:factorization},
which is in agreement with the discrepancy observed in figure~\ref{fig:compare_vnpt_lowptass}.

The $\pt$-averaged $v_n$ values (with \ptref\ of 1--3\GeVc) weighted by the
efficiency-corrected charged-hadron yield, over the \pt range from 0.3 to 3.0\GeVc,
are shown in Fig.~\ref{fig:compare_vnn_centrality} as a function of $n$ up to $n=7$
(the $v_7$ value as a function of \pt is not presented in Fig.~\ref{fig:compare_vnpt_ucc}
due to limited statistical precision). The 0--0.2\% ultra-central events are compared
to several other very central \PbPb\ centrality ranges including 2.5--5.0\%, 0--2.5\%,
0--1\% and 0--0.02\%. As mentioned earlier, results for 0--1\% centrality are compared
with both the HF \ET\ sum selection (not shown) and HF \ET\ sum plus pixel cluster
multiplicity (NPixel) selection as a systematic check. The two methods of centrality selection
yield consistent $v_n$ results within statistical uncertainties. Therefore, only results
from HF \ET\ sum plus pixel cluster multiplicity centrality selection are shown
in Fig.~\ref{fig:compare_vnn_centrality}. Beyond the 2.5--5.0\% centrality range,
the $v_n$ values are still decreasing toward more central collisions, especially
for $v_2$. Going from 0--0.2\% to 0--0.02\% centrality, $v_n$ shows almost no change,
indicating events do not become significantly more central by requiring larger HF
\ET\ sum and pixel cluster multiplicity, especially in terms of
eccentricities. This is consistent with the studies using the \textsc{ampt} model. The $v_n$
values remain finite up to $n=6$ within the statistical precision of our data.
Beyond $n=6$, $v_n$ becomes consistent with zero. The magnitude of $v_2$ and $v_3$
are very similar, while the $v_n$ become progressively smaller for $n \geq 4$.
This is qualitatively in agreement with expectations from hydrodynamic
calculations~\cite{Alver:2010dn}.

\subsection{Correlation Functions}

\begin{figure}[thb]
  \begin{center}
    \includegraphics[width=\linewidth]{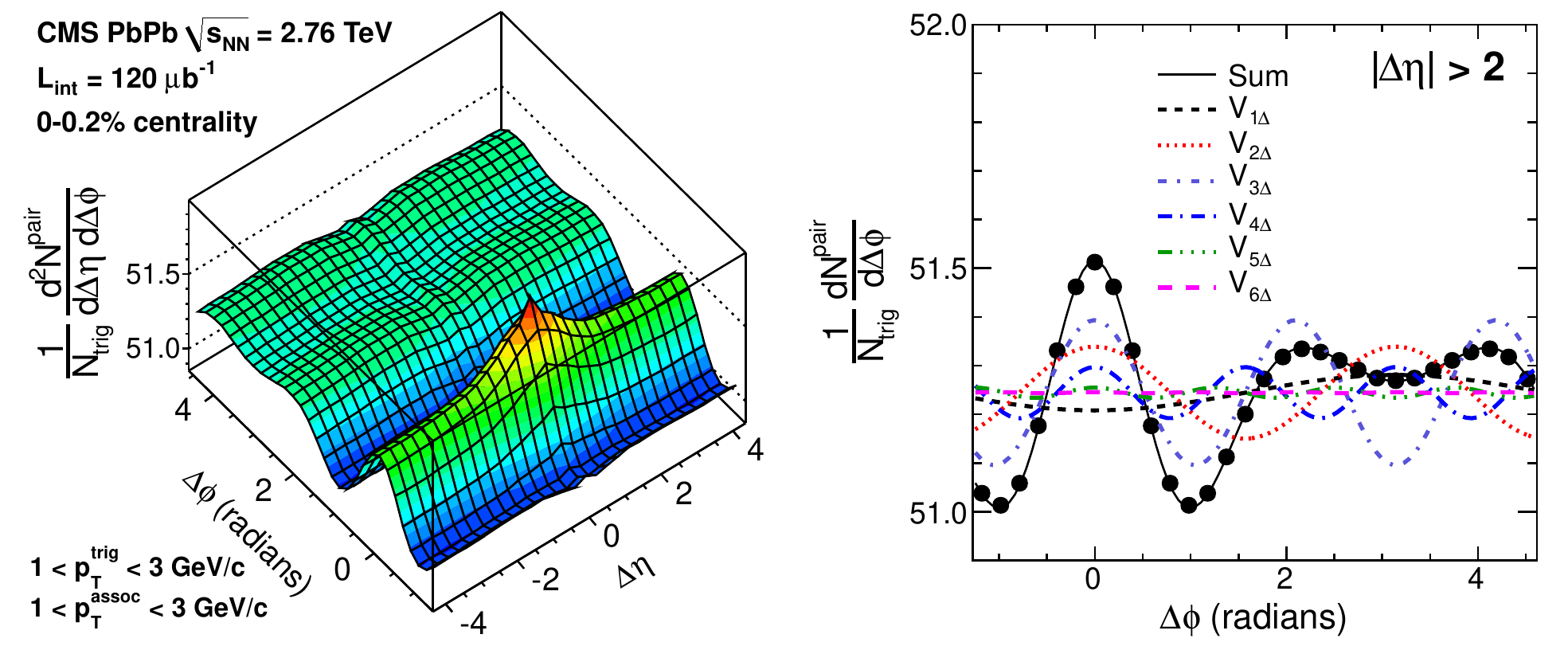}
    \caption{The 2D (left) and 1D \dphi\ (right) dihadron correlation functions for
    $1<\pttrg<3$\GeVc and $1<\ptass<3$\GeVc in 0--0.2\% central \PbPb\ collisions at
    \rootsNN\ = 2.76\TeV. The broken lines on the right panel show various orders of
    $V_{n\Delta}$ components expected from the extracted $v_n$ values in Section~\ref{sec:vn}, while
    the solid line is the sum of all $V_{n\Delta}$ components.}
    \label{fig:UCC_corrfunc}
  \end{center}
\end{figure}

Dihadron correlation functions are also constructed using Eq.~(\ref{2pcorr_incl})
in order to check the consistency of extracting $V_{n\Delta}$ using Eq.~(\ref{average})
with the fit method to the correlation function by a Fourier series in Eq.~(\ref{fourier}).
Figure~\ref{fig:UCC_corrfunc} (left) shows the dihadron
correlation functions for $1<\pttrg<3$\GeVc and $1<\ptass<3$\GeVc in 0--0.2\%
central \PbPb\ collisions at \rootsNN\ = 2.76\TeV. As shown in Fig.~\ref{fig:compare_vnpt_ucc},
the $v_3$, $v_4$, and $v_5$ values become comparable or even bigger than $v_2$ at $1<\pt<3$\GeVc.
In Fig.~\ref{fig:UCC_corrfunc}, this can be seen in the dihadron correlation
function on the away side ($\dphi \sim \pi$), where a significant local minimum
(at $\dphi \sim \pi$ along $\deta$) is present.
On the near side ($\dphi \sim 0$) of the correlation function, a long-range structure
extending over the entire \deta\ region is present. The observed features of
the correlation function are similar to what was seen previously at CMS in other
centrality ranges of \PbPb\ collisions~\cite{Chatrchyan:2011eka,Chatrchyan:2012wg},
although the dip on the away side is not seen in non-central \PbPb\ collisions. This
may indicate that the contribution of higher-order Fourier components (e.g., $v_3$)
is more relevant for very central events.

Averaging over $\Delta\eta$, the 1D $\Delta\phi$ dihadron correlation
function, for $1<\pttrg<3$\GeVc and $1<\ptass<3$\GeVc in 0--0.2\% central \PbPb\ collisions at \rootsNN\ = 2.76\TeV, is shown in Fig.~\ref{fig:UCC_corrfunc} (right).
The range of $|\Delta\eta|<2$ is excluded from the average to avoid
non-flow effects from other source of correlations, such as jet fragmentation.
The dashed curves represent different $V_{n\Delta}$
components and are constructed from the $v_n$ values extracted in Section~\ref{sec:vn}
by assuming factorization. The solid curve is the sum of all $V_{n\Delta}$
components, which is in good agreement with the measured dihadron correlation function.

\subsection{Factorization breakdown and \texorpdfstring{\pt}{pt} dependence of event plane angle}
\label{sec:factorization}

The breakdown of factorization observed in Fig.~\ref{fig:compare_vnpt_lowptass}
could be caused by non-flow effects that contribute to the dihadron
correlation function at large $\deta$, e.g., back-to-back jet correlations.
However, in hydrodynamics, it has been recently suggested that one possible source of
factorization breakdown is related to the initial-state eccentricity
fluctuations~\cite{Gardim:2012im,Heinz}. The event plane angle, $\Psi_n$,
as determined by final-state particles, could be dependent on the particle
\pt event-by-event, instead of a unique angle for the entire event (which
is the case for a non-fluctuating smooth initial condition). Because of this
effect, the factorization of $V_{n\Delta}$ extracted from dihadron correlations
could be broken, even if hydrodynamic flow is the only source of correlations.
The breakdown effect can be explored more quantitatively in the following analysis.

A ratio for testing factorization defined as
\begin{equation}
    \label{r_n_def}
    r_{n} \equiv \frac{V_{n\Delta}(\pttrg,\ptass)}{\sqrt{V_{n\Delta}(\pttrg,\pttrg)V_{n\Delta}(\ptass,\ptass)}}
\end{equation}
has been proposed as a direct measurement of $\pt$-dependent event plane angle fluctuations~\cite{Heinz}.
Here, the $V_{n\Delta}$ coefficients are calculated by pairing particles
within the same $\pt$ interval (denominator) or from different $\pt$
intervals (numerator). If $V_{n\Delta}$ factorizes, this ratio will be equal to
unity. With the presence of a $\pt$-dependent event plane angle,
it has been shown that the ratio, $r_n$, is equivalent to
\begin{equation}
    \label{r_n}
    r_{n} =\frac{\langle v_{n}(\pttrg)v_{n}(\ptass)\cos\big[n\big(\Psi_{n}(\pttrg)-\Psi_{n}(\ptass)\big)\big] \rangle}{\sqrt{\langle v_{n}^{2}(\pttrg) \rangle\langle v_{n}^{2}(\ptass) \rangle}},
\end{equation}
where $\Psi_{n}(\pttrg)$ and $\Psi_{n}(\ptass)$ represent
the event plane angles determined for trigger and associated particles
from two \pt intervals~\cite{Gardim:2012im,Heinz}.
One can see from Eq.~(\ref{r_n}) that $r_n$ is in general
less than unity if event plane angle $\Psi_{n}$ depends on $\pt$.

\begin{figure}[thb]
  \begin{center}
    \includegraphics[width=\linewidth]{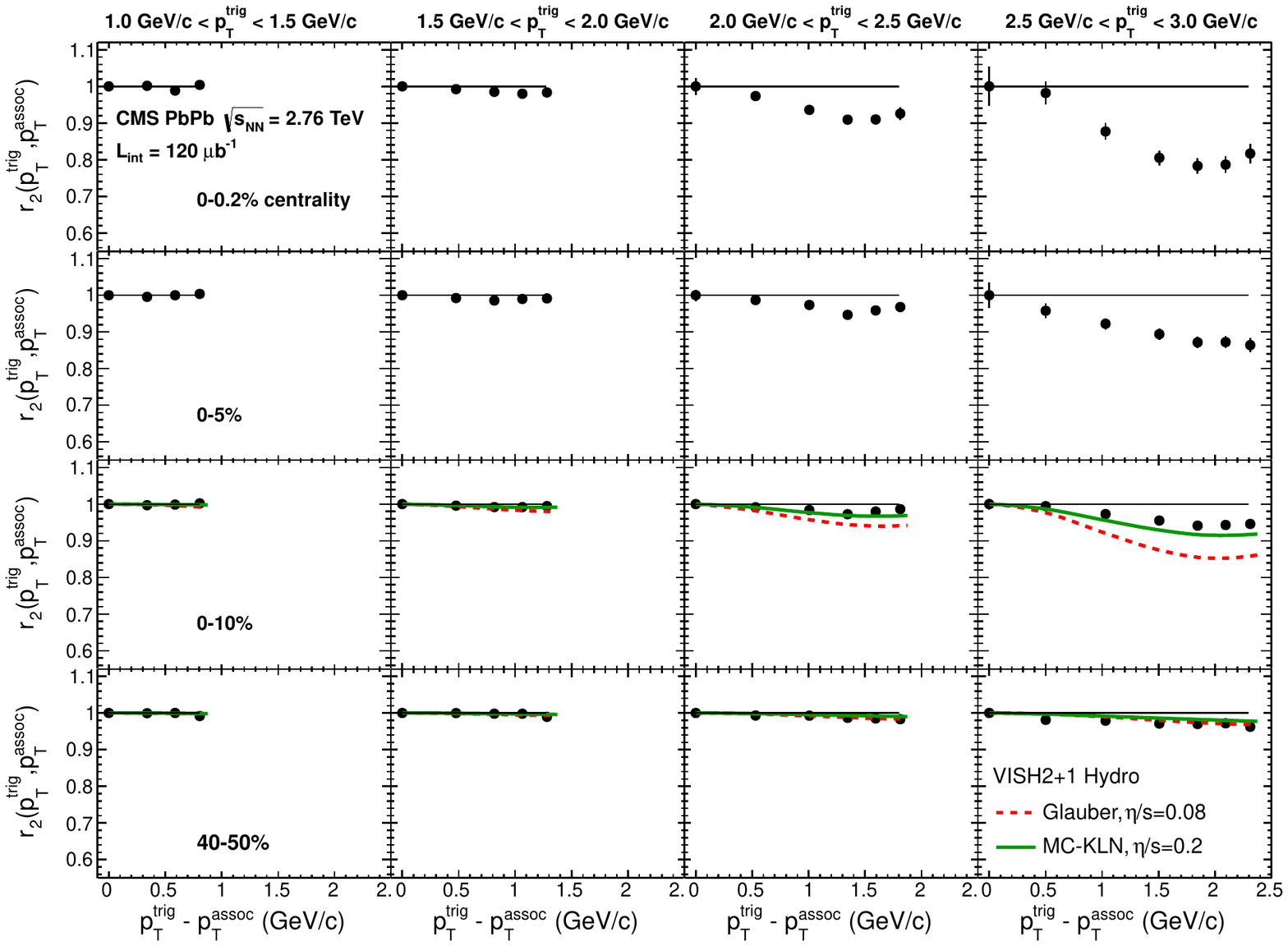}
    \caption{Factorization ratio, $r_2$, as a function of \pttrg\ - \ptass\ in bins
    of \pttrg\ for four centrality ranges of \PbPb\ collisions at \rootsNN\ = 2.76\TeV.
    The lines show the calculations from viscous hydrodynamics in Ref.~\cite{Heinz}
    for 0--10\% and 40--50\% centralities with MC Glauber initial condition model
    and $\eta/s=0.08$ (dashed lines), and MC-KLN initial condition model and $\eta/s=0.2$ (solid lines).
    Each row represents a different centrality range, while each column corresponds to a
    different \pttrg\ range. The error bars correspond to statistical uncertainties,
    while systematic uncertainties are negligible for the $r_n$ ratios, and thus are
    not presented.
    }
    \label{fig:r2}
  \end{center}
\end{figure}

\begin{figure}[thb]
  \begin{center}
    \includegraphics[width=\linewidth]{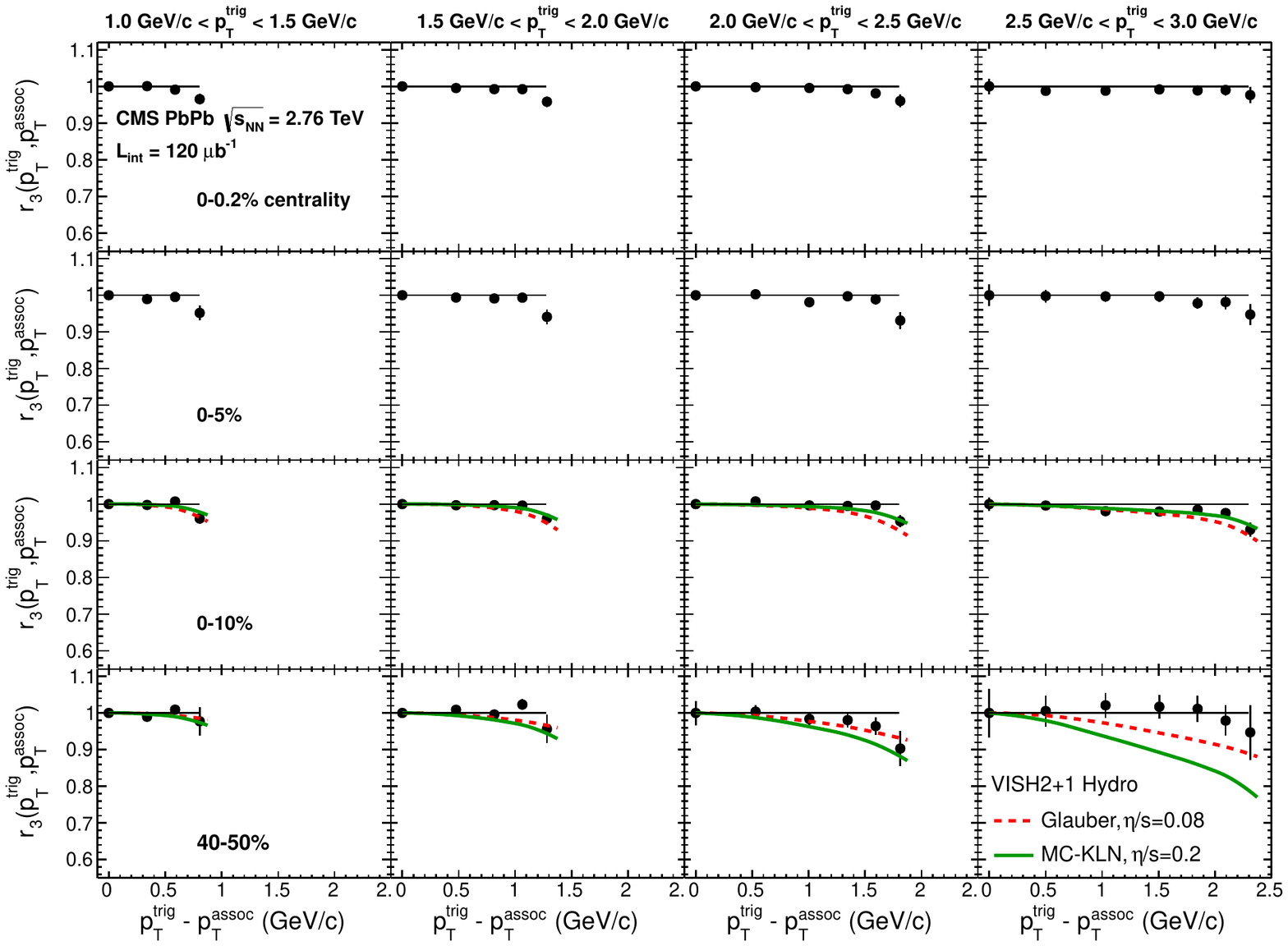}
    \caption{Factorization ratio, $r_3$, as a function of \pttrg\ - \ptass\ in bins
    of \pttrg\ for four centrality ranges of \PbPb\ collisions at \rootsNN\ = 2.76\TeV.
    The lines show the calculations from viscous hydrodynamics in Ref.~\cite{Heinz}
    for 0--10\% and 40--50\% centralities with MC Glauber initial condition model
    and $\eta/s=0.08$ (dashed lines), and MC-KLN initial condition model and $\eta/s=0.2$ (solid lines).
    Each row represents a different centrality range, while each column corresponds to a
    different \pttrg\ range. The error bars correspond to statistical uncertainties,
    while systematic uncertainties are negligible for the $r_n$ ratios, and thus are
    not presented.
    }
    \label{fig:r3}
  \end{center}
\end{figure}

\begin{figure}[thb]
  \begin{center}
    \includegraphics[width=\linewidth]{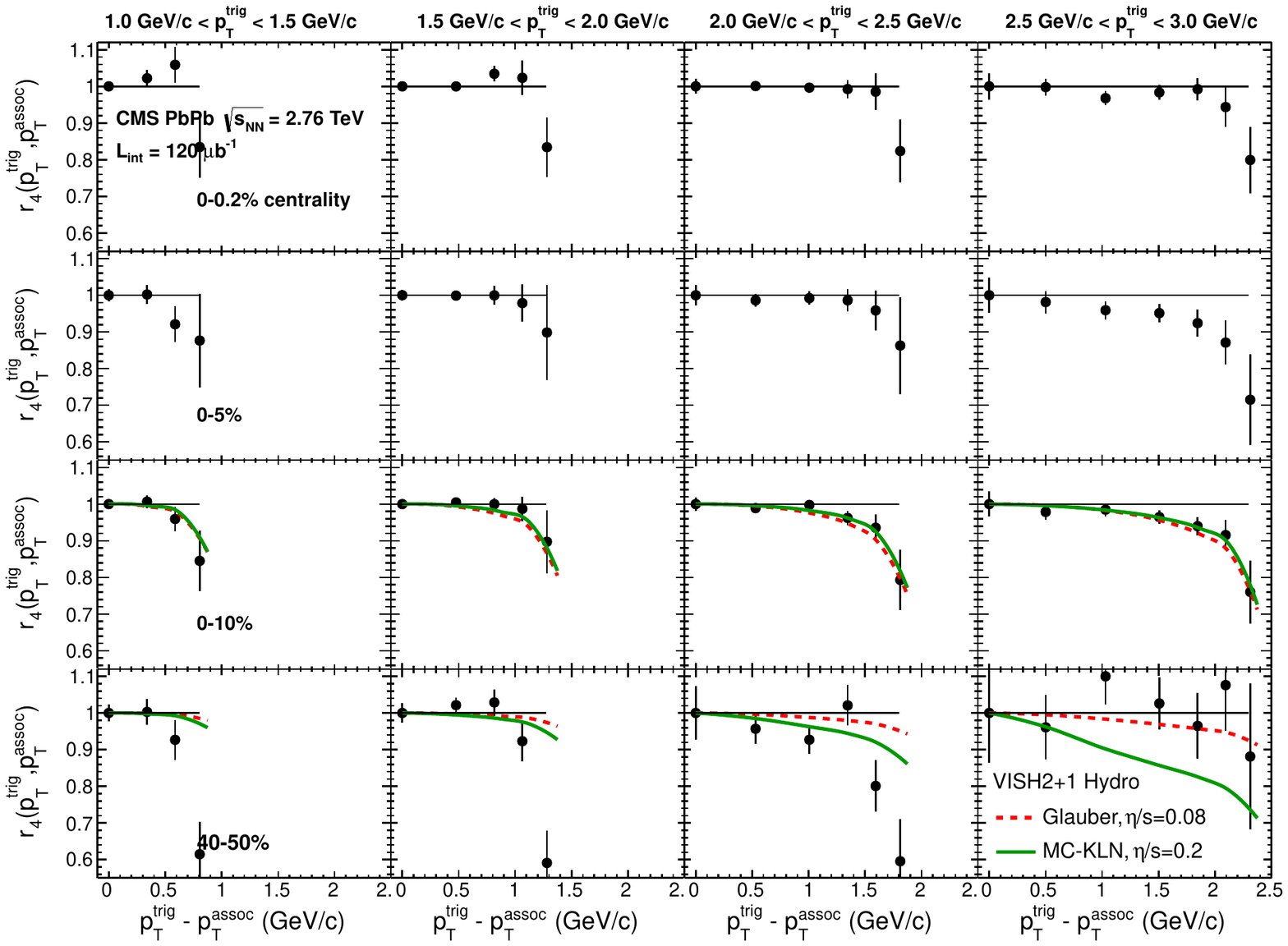}
    \caption{Factorization ratio, $r_4$, as a function of \pttrg\ - \ptass\ in bins
    of \pttrg\ for four centrality ranges of \PbPb\ collisions at \rootsNN\ = 2.76\TeV.
    The lines show the calculations from viscous hydrodynamics in Ref.~\cite{Heinz}
    for 0--10\% and 40--50\% centralities with MC Glauber initial condition model
    and $\eta/s=0.08$ (dashed lines), and MC-KLN initial condition model and $\eta/s=0.2$ (solid lines).
    Each row represents a different centrality range, while each column corresponds to a
    different \pttrg\ range. The error bars correspond to statistical uncertainties,
    while systematic uncertainties are negligible for the $r_n$ ratios, and thus are
    not presented.
    }
    \label{fig:r4}
  \end{center}
\end{figure}

In this paper, the proposed factorization ratio, $r_{n}$, is studied
as a function of \pttrg\ and \ptass\ for different centrality classes
in \PbPb\ collisions at \rootsNN\ = 2.76\TeV. Figures~\ref{fig:r2}--\ref{fig:r4}
show the $r_n$ values for $n=2$--$4$, respectively, for four \pttrg\ bins
(of increasing \pt from left to right panels) as a function of the difference between
\pttrg\ and \ptass. The average values of \pttrg\ and \ptass\ in each bin
are used for calculating the difference. The measurement
is performed in four different centrality classes, i.e., 40--50\%, 0--10\%,
0--5\%, and ultra-central 0--0.2\% centralities (from bottom to top panels).
By construction, the $r_{n}$ value for the highest analyzed \ptass\ range,
where trigger and associated particles are selected from the same \pt
interval, is equal to one. Only results for \pttrg\ $\geq$ \ptass\ are presented.
The error bars correspond to statistical uncertainties, while systematic
uncertainties are negligible for the $r_n$ ratios, and thus are not presented
in the figures.

For the second Fourier harmonics (Fig.~\ref{fig:r2}), the $r_2$ ratio
significantly deviates from one as the collisions become more central.
For any centrality, the effect gets larger with an increase of the difference
between \pttrg\ and \ptass\ values. To explicitly emphasize this observation,
$\pttrg - \ptass$, instead of \ptass, is used as the horizontal axis of
figures~\ref{fig:r2}--\ref{fig:r4}. The deviation reaches up to 20\%
for the lowest \ptass\ bins in the ultra-central 0--0.2\% events for $2.5<\pttrg<3.0$.
This is expected as event-by-event initial-state geometry fluctuations play a more dominant role
as the collisions become more central. Calculations from viscous hydrodynamics in
Ref.~\cite{Heinz} are compared to data for 0--10\% and 40--50\% centralities
with MC Glauber initial condition model~\cite{glauber,Alver:Glauber}
and $\eta/s=0.08$ (dashed lines), and MC-KLN initial condition
model~\cite{Drescher:2006pi} and $\eta/s=0.2$ (solid lines). The qualitative trend
of hydrodynamic calculations is the same as what is observed in the data.
The observed $r_2$ values are found to be more consistent with the MC-KLN model and
an $\eta/s$ value of 0.2. However, future theoretical studies, particularly with comparison
to the precision ultra-central collisions data presented in this paper, are still
needed to achieve better constraints on the initial-state models and the $\eta/s$ value
of the system.

For higher-order harmonics ($n=3,4$), shown in Fig.~\ref{fig:r3} and Fig.~\ref{fig:r4},
the factorization is fulfilled over a wider range of \pttrg, \ptass, and centrality ranges
than for $v_2$. The factorization only breaks by about 5\% at large values of $\pttrg - \ptass$, i.e.,
greater than 1\GeVc. Due to large statistical uncertainties, $r_5$ is not included in this result.
Again, the qualitative trend of the data is described by hydrodynamics for 0--10\% centrality,
while no conclusion can be drawn for 40--50\% centrality based on the present
statistical precision of the data.

\section{Conclusion}

In summary, azimuthal dihadron correlations were studied for \PbPb\ collisions
at \rootsNN\ = 2.76\TeV using the CMS detector at the LHC.
Assuming factorization, these two-particle correlations were
used to extract the single-particle anisotropy harmonics, $v_n$,
as a function of \pt from 0.3 to 8.0\GeVc. The data set includes a
sample of ultra-central (0--0.2\% centrality) \PbPb\ events collected
using a trigger based on total transverse energy in the hadron forward
calorimeters and the total multiplicity of pixel clusters in the silicon
pixel tracker. In the context of
hydrodynamic models, anisotropies in such ultra-central heavy-ion
collisions arise predominantly from initial-state eccentricity
fluctuations. The magnitude of the flow harmonics decreases from
$v_3$ to $v_6$. As a function of \pt, these four harmonics all display
a common maximum around $\pt = 3.5$\GeVc. Although
the $v_2$ harmonic exceeds the others at low \pt, it falls below $v_3$
around $\pt = 1$\GeVc and reaches its maximum around $\pt = 2.5$\GeVc.

The $\pt$-averaged $v_n$ for $0.3<\pt<3.0$\GeVc were also
derived up to $n=7$, and results for 0--0.2\% collisions were compared
to those for other slightly less central ranges. Between the
2.5--5.0\% and 0--0.2\% centrality ranges, all $v_n$ harmonics
decrease. The decrease is largest for $v_2$, reaching up to 45\%.
Only small variations of $v_n$ are observed for events that are even
more central than 0--0.2\% (e.g., 0--0.02\%).
For the most central collisions, the $\pt$-averaged
$v_2$ and $v_3$ are found to be comparable within 2\%, while
higher-order $v_n$ decrease as $n$ increases.

Detailed studies indicate that factorization of dihadron correlations
into single-particle azimuthal anisotropies does not hold precisely.
The observed breakdown of factorization increases up to about 20\%
as the \pt difference between the two particles becomes larger in ultra-central
\PbPb\ events. This behavior is expected in hydrodynamic models,
in which a $\pt$-dependent event plane angle is induced by
initial-state fluctuations. The factorization data for the 0--10\% and
40--50\% centrality ranges were compared to viscous hydrodynamic calculations
with different models of initial-state fluctuations and different $\eta/s$ values.
Future quantitative theoretical
comparisons to the high-precision data of ultra-central \PbPb\ collisions
presented by the CMS collaboration in this paper can provide a new stringent
test of hydrodynamic models, particularly for constraining the initial-state
density fluctuations and the $\eta/s$ value.

\section{Acknowledgment}

We congratulate our colleagues in the CERN accelerator departments for the excellent
performance of the LHC and thank the technical and administrative staffs at CERN and
at other CMS institutes for their contributions to the success of the CMS effort. In
addition, we gratefully acknowledge the computing centres and personnel of the Worldwide
LHC Computing Grid for delivering so effectively the computing infrastructure essential
to our analyses. Finally, we acknowledge the enduring support for the construction and
operation of the LHC and the CMS detector provided by the following funding agencies:
BMWF and FWF (Austria); FNRS and FWO (Belgium); CNPq, CAPES, FAPERJ, and FAPESP (Brazil);
MES (Bulgaria); CERN; CAS, MoST, and NSFC (China); COLCIENCIAS (Colombia); MSES (Croatia);
RPF (Cyprus); MoER, SF0690030s09 and ERDF (Estonia); Academy of Finland, MEC, and HIP
(Finland); CEA and CNRS/IN2P3 (France); BMBF, DFG, and HGF (Germany); GSRT (Greece);
OTKA and NKTH (Hungary); DAE and DST (India); IPM (Iran); SFI (Ireland); INFN (Italy);
NRF and WCU (Republic of Korea); LAS (Lithuania); CINVESTAV, CONACYT, SEP, and UASLP-FAI
(Mexico); MBIE (New Zealand); PAEC (Pakistan); MSHE and NSC (Poland); FCT (Portugal);
JINR (Dubna); MON, RosAtom, RAS and RFBR (Russia); MESTD (Serbia); SEIDI and CPAN (Spain);
Swiss Funding Agencies (Switzerland); NSC (Taipei); ThEPCenter, IPST, STAR and NSTDA
(Thailand); TUBITAK and TAEK (Turkey); NASU (Ukraine); STFC (United Kingdom); DOE and NSF (USA).

Individuals have received support from the Marie-Curie programme and the European Research
Council and EPLANET (European Union); the Leventis Foundation; the A. P. Sloan Foundation;
the Alexander von Humboldt Foundation; the Belgian Federal Science Policy Office; the Fonds
pour la Formation \`a la Recherche dans l'Industrie et dans l'Agriculture (FRIA-Belgium);
the Agentschap voor Innovatie door Wetenschap en Technologie (IWT-Belgium); the Ministry
of Education, Youth and Sports (MEYS) of Czech Republic; the Council of Science and Industrial
Research, India; the Compagnia di San Paolo (Torino); the HOMING PLUS programme of Foundation
for Polish Science, cofinanced by EU, Regional Development Fund; and the Thalis and Aristeia
programmes cofinanced by EU-ESF and the Greek NSRF.

\clearpage
\bibliography{auto_generated}   

\cleardoublepage \appendix\section{The CMS Collaboration \label{app:collab}}\begin{sloppypar}\hyphenpenalty=5000\widowpenalty=500\clubpenalty=5000\input{HIN-12-011-authorlist.tex}\end{sloppypar}
\end{document}

%% file: HIN-12-011-authorlist.tex
\textbf{Yerevan Physics Institute,  Yerevan,  Armenia}\\*[0pt]
S.~Chatrchyan, V.~Khachatryan, A.M.~Sirunyan, A.~Tumasyan
\vskip\cmsinstskip
\textbf{Institut f\"{u}r Hochenergiephysik der OeAW,  Wien,  Austria}\\*[0pt]
W.~Adam, T.~Bergauer, M.~Dragicevic, J.~Er\"{o}, C.~Fabjan\cmsAuthorMark{1}, M.~Friedl, R.~Fr\"{u}hwirth\cmsAuthorMark{1}, V.M.~Ghete, C.~Hartl, N.~H\"{o}rmann, J.~Hrubec, M.~Jeitler\cmsAuthorMark{1}, W.~Kiesenhofer, V.~Kn\"{u}nz, M.~Krammer\cmsAuthorMark{1}, I.~Kr\"{a}tschmer, D.~Liko, I.~Mikulec, D.~Rabady\cmsAuthorMark{2}, B.~Rahbaran, H.~Rohringer, R.~Sch\"{o}fbeck, J.~Strauss, A.~Taurok, W.~Treberer-Treberspurg, W.~Waltenberger, C.-E.~Wulz\cmsAuthorMark{1}
% ...
\vskip\cmsinstskip
% ...
1:~~Also at Vienna University of Technology, Vienna, Austria\\
2:~~Also at CERN, European Organization for Nuclear Research, Geneva, Switzerland\\
% ...


%% file: HIN-12-011_temp.bbl
\providecommand{\href}[2]{#2}\begingroup\raggedright\begin{thebibliography}{10}%
\makeatletter
\providecommand{\hrefCMSnoop }[0]{\@secondoftwo}%
\makeatother
\providecommand{\doi}{\texttt{doi:}\begingroup \urlstyle{tt}\Url}

\bibitem{Heinz:2013th}
\hrefCMSnoop {} {U.~Heinz and R.~Snellings, ``{Collective flow and viscosity in
  relativistic heavy-ion collisions}'',} \textit{ Ann. Rev. Nucl. Part. Sci.}
  \textbf{ 63} (2013) 123,
  \href{http://dx.doi.org/10.1146/annurev-nucl-102212-170540}{\doi{10.1146/annurev-nucl-102212-170540}},
\href{http://www.arXiv.org/abs/1301.2826}{\texttt{ arXiv:1301.2826}}.

\bibitem{PHENIX}
\hrefCMSnoop {} {{ PHENIX} Collaboration, ``{Formation of dense partonic matter
  in relativistic nucleus-nucleus collisions at RHIC: Experimental evaluation
  by the PHENIX collaboration}'',} \textit{ Nucl. Phys. A} \textbf{ 757} (2005)
  184,
  \href{http://dx.doi.org/10.1016/j.nuclphysa.2005.03.086}{\doi{10.1016/j.nuclphysa.2005.03.086}},
\href{http://www.arXiv.org/abs/nucl-ex/0410003}{\texttt{
  arXiv:nucl-ex/0410003}}.

\bibitem{STAR}
\hrefCMSnoop {} {{ STAR} Collaboration, ``{Experimental and theoretical
  challenges in the search for the quark gluon plasma: The STAR Collaboration's
  critical assessment of the evidence from RHIC collisions}'',} \textit{ Nucl.
  Phys. A} \textbf{ 757} (2005) 102,
  \href{http://dx.doi.org/10.1016/j.nuclphysa.2005.03.085}{\doi{10.1016/j.nuclphysa.2005.03.085}},
\href{http://www.arXiv.org/abs/nucl-ex/0501009}{\texttt{
  arXiv:nucl-ex/0501009}}.

\bibitem{PHOBOS}
\hrefCMSnoop {} {{ PHOBOS} Collaboration, ``{The PHOBOS perspective on
  discoveries at RHIC}'',} \textit{ Nucl. Phys. A} \textbf{ 757} (2005) 28,
  \href{http://dx.doi.org/10.1016/j.nuclphysa.2005.03.084}{\doi{10.1016/j.nuclphysa.2005.03.084}},
\href{http://www.arXiv.org/abs/nucl-ex/0410022}{\texttt{
  arXiv:nucl-ex/0410022}}.

\bibitem{BRAHMS}
\hrefCMSnoop {} {{ BRAHMS} Collaboration, ``{Quark gluon plasma and color glass
  condensate at RHIC? The Perspective from the BRAHMS experiment}'',} \textit{
  Nucl. Phys. A} \textbf{ 757} (2005) 1,
  \href{http://dx.doi.org/10.1016/j.nuclphysa.2005.02.130}{\doi{10.1016/j.nuclphysa.2005.02.130}},
\href{http://www.arXiv.org/abs/nucl-ex/0410020}{\texttt{
  arXiv:nucl-ex/0410020}}.

\bibitem{Shuryak:2004cy}
\hrefCMSnoop {} {E.~V. Shuryak, ``{What RHIC experiments and theory tell us
  about properties of quark-gluon plasma?}'',} \textit{ Nucl. Phys. A} \textbf{
  750} (2005) 64,
  \href{http://dx.doi.org/10.1016/j.nuclphysa.2004.10.022}{\doi{10.1016/j.nuclphysa.2004.10.022}},
\href{http://www.arXiv.org/abs/hep-ph/0405066}{\texttt{ arXiv:hep-ph/0405066}}.

\bibitem{Gyulassy:2004zy}
\hrefCMSnoop {} {M.~Gyulassy and L.~McLerran, ``{New forms of QCD matter
  discovered at RHIC}'',} \textit{ Nucl. Phys. A} \textbf{ 750} (2005) 30,
  \href{http://dx.doi.org/10.1016/j.nuclphysa.2004.10.034}{\doi{10.1016/j.nuclphysa.2004.10.034}},
\href{http://www.arXiv.org/abs/nucl-th/0405013}{\texttt{
  arXiv:nucl-th/0405013}}.

\bibitem{Chatrchyan:2012ta}
\hrefCMSnoop {} {{ CMS} Collaboration, ``{Measurement of the elliptic
  anisotropy of charged particles produced in PbPb collisions at
  nucleon-nucleon center-of-mass energy = 2.76 TeV}'',} \textit{ Phys. Rev. C}
  \textbf{ 87} (2013) 014902,
  \href{http://dx.doi.org/10.1103/PhysRevC.87.014902}{\doi{10.1103/PhysRevC.87.014902}},
\href{http://www.arXiv.org/abs/1204.1409}{\texttt{ arXiv:1204.1409}}.

\bibitem{Chatrchyan:2012wg}
\hrefCMSnoop {} {{ CMS} Collaboration, ``{Centrality dependence of dihadron
  correlations and azimuthal anisotropy harmonics in PbPb collisions at
  \rootsNN\ = 2.76 TeV}'',} \textit{ Eur. Phys. J. C} \textbf{ 72} (2012) 2012,
  \href{http://dx.doi.org/10.1140/epjc/s10052-012-2012-3}{\doi{10.1140/epjc/s10052-012-2012-3}},
\href{http://www.arXiv.org/abs/1201.3158}{\texttt{ arXiv:1201.3158}}.

\bibitem{Chatrchyan:2012xq}
\hrefCMSnoop {} {{ CMS} Collaboration, ``{Azimuthal anisotropy of charged
  particles at high transverse momenta in PbPb collisions at \rootsNN\ =
  2.76\TeV}'',} \textit{ Phys. Rev. Lett.} \textbf{ 109} (2012) 022301,
  \href{http://dx.doi.org/10.1103/PhysRevLett.109.022301}{\doi{10.1103/PhysRevLett.109.022301}},
\href{http://www.arXiv.org/abs/1204.1850}{\texttt{ arXiv:1204.1850}}.

\bibitem{Chatrchyan:2012vqa}
\hrefCMSnoop {} {{ CMS} Collaboration, ``{Measurement of the azimuthal
  anisotropy of neutral pions in PbPb collisions at \rootsNN\ = 2.76\TeV}'',}
  \textit{ Phys. Rev. Lett.} \textbf{ 110} (2013) 042301,
  \href{http://dx.doi.org/10.1103/PhysRevLett.110.042301}{\doi{10.1103/PhysRevLett.110.042301}},
\href{http://www.arXiv.org/abs/1208.2470}{\texttt{ arXiv:1208.2470}}.

\bibitem{Aamodt:2010pa}
\hrefCMSnoop {} {{ ALICE} Collaboration, ``{Elliptic flow of charged particles
  in Pb-Pb collisions at 2.76 TeV}'',} \textit{ Phys. Rev. Lett.} \textbf{ 105}
  (2010) 252302,
  \href{http://dx.doi.org/10.1103/PhysRevLett.105.252302}{\doi{10.1103/PhysRevLett.105.252302}},
\href{http://www.arXiv.org/abs/1011.3914}{\texttt{ arXiv:1011.3914}}.

\bibitem{ALICE:2011ab}
\hrefCMSnoop {} {{ ALICE} Collaboration, ``{Higher harmonic anisotropic flow
  measurements of charged particles in Pb-Pb collisions at \rootsNN\ = 2.76
  TeV}'',} \textit{ Phys. Rev. Lett.} \textbf{ 107} (2011) 032301,
  \href{http://dx.doi.org/10.1103/PhysRevLett.107.032301}{\doi{10.1103/PhysRevLett.107.032301}},
\href{http://www.arXiv.org/abs/1105.3865}{\texttt{ arXiv:1105.3865}}.

\bibitem{Aamodt:2011by}
\hrefCMSnoop {} {{ ALICE} Collaboration, ``{Harmonic decomposition of
  two-particle angular correlations in Pb-Pb collisions at \rootsNN\ = 2.76
  TeV}'',} \textit{ Phys. Lett. B} \textbf{ 708} (2012) 249,
  \href{http://dx.doi.org/10.1016/j.physletb.2012.01.060}{\doi{10.1016/j.physletb.2012.01.060}},
\href{http://www.arXiv.org/abs/1109.2501}{\texttt{ arXiv:1109.2501}}.

\bibitem{ATLAS:2011ah}
\hrefCMSnoop {} {{ ATLAS} Collaboration, ``{Measurement of the pseudorapidity
  and transverse momentum dependence of the elliptic flow of charged particles
  in lead-lead collisions at \rootsNN\ = 2.76 TeV with the ATLAS detector}'',}
  \textit{ Phys. Lett. B} \textbf{ 707} (2012) 330,
  \href{http://dx.doi.org/10.1016/j.physletb.2011.12.056}{\doi{10.1016/j.physletb.2011.12.056}},
\href{http://www.arXiv.org/abs/1108.6018}{\texttt{ arXiv:1108.6018}}.

\bibitem{ATLAS:2012at}
\hrefCMSnoop {} {{ ATLAS} Collaboration, ``{Measurement of the azimuthal
  anisotropy for charged particle production in \rootsNN\ = 2.76\TeV lead-lead
  collisions with the ATLAS detector}'',} \textit{ Phys. Rev. C} \textbf{ 86}
  (2012) 014907,
  \href{http://dx.doi.org/10.1103/PhysRevC.86.014907}{\doi{10.1103/PhysRevC.86.014907}},
\href{http://www.arXiv.org/abs/1203.3087}{\texttt{ arXiv:1203.3087}}.

\bibitem{Aad:2013xma}
\hrefCMSnoop {} {{ ATLAS} Collaboration, ``{Measurement of the distributions of
  event-by-event flow harmonics in lead-lead collisions at \rootsNN\ = 2.76\TeV
  with the ATLAS detector at the LHC}'',} \textit{ JHEP} \textbf{ 11} (2013)
  183,
  \href{http://dx.doi.org/10.1007/JHEP11(2013)183}{\doi{10.1007/JHEP11(2013)183}},
\href{http://www.arXiv.org/abs/1305.2942}{\texttt{ arXiv:1305.2942}}.

\bibitem{PhysRevLett.104.142301}
\hrefCMSnoop {} {{ PHOBOS} Collaboration, ``{Event-by-Event Fluctuations of
  Azimuthal Particle Anisotropy in Au + Au Collisions at \rootsNN\ =
  200\GeV}'',} \textit{ Phys. Rev. Lett.} \textbf{ 104} (2010) 142301,
  \href{http://dx.doi.org/10.1103/PhysRevLett.104.142301}{\doi{10.1103/PhysRevLett.104.142301}},
\href{http://www.arXiv.org/abs/nucl-ex/0702036}{\texttt{
  arXiv:nucl-ex/0702036}}.

\bibitem{PHOBOSeccPART}
\hrefCMSnoop {} {B.~Alver {et~al.}, ``{Importance of correlations and
  fluctuations on the initial source eccentricity in high-energy
  nucleus-nucleus collisions}'',} \textit{ Phys. Rev. C} \textbf{ 77} (2008)
  014906,
  \href{http://dx.doi.org/10.1103/PhysRevC.77.014906}{\doi{10.1103/PhysRevC.77.014906}},
\href{http://www.arXiv.org/abs/0711.3724}{\texttt{ arXiv:0711.3724}}.

\bibitem{Bhalerao:2006tp}
\hrefCMSnoop {} {R.~S. Bhalerao and J.-Y. Ollitrault, ``{Eccentricity
  fluctuations and elliptic flow at RHIC}'',} \textit{ Phys. Lett. B} \textbf{
  641} (2006) 260,
  \href{http://dx.doi.org/10.1016/j.physletb.2006.08.055}{\doi{10.1016/j.physletb.2006.08.055}},
\href{http://www.arXiv.org/abs/nucl-th/0607009}{\texttt{
  arXiv:nucl-th/0607009}}.

\bibitem{Voloshin:2007pc}
\hrefCMSnoop {} {S.~A. Voloshin, A.~M. Poskanzer, A.~Tang, and G.~Wang,
  ``{Elliptic flow in the Gaussian model of eccentricity fluctuations}'',}
  \textit{ Phys. Lett. B} \textbf{ 659} (2008) 537,
  \href{http://dx.doi.org/10.1016/j.physletb.2007.11.043}{\doi{10.1016/j.physletb.2007.11.043}},
\href{http://www.arXiv.org/abs/0708.0800}{\texttt{ arXiv:0708.0800}}.

\bibitem{Ollitrault:2009ie}
\hrefCMSnoop {} {J.-Y. Ollitrault, A.~M. Poskanzer, and S.~A. Voloshin,
  ``{Effect of flow fluctuations and nonflow on elliptic flow methods}'',}
  \textit{ Phys. Rev. C} \textbf{ 80} (2009) 014904,
  \href{http://dx.doi.org/10.1103/PhysRevC.80.014904}{\doi{10.1103/PhysRevC.80.014904}},
\href{http://www.arXiv.org/abs/0904.2315}{\texttt{ arXiv:0904.2315}}.

\bibitem{Alver:2010gr}
\hrefCMSnoop {} {B.~Alver and G.~Roland, ``{Collision geometry fluctuations and
  triangular flow in heavy-ion collisions}'',} \textit{ Phys. Rev. C} \textbf{
  81} (2010) 054905,
  \href{http://dx.doi.org/10.1103/PhysRevC.81.054905}{\doi{10.1103/PhysRevC.81.054905}},
  \href{http://www.arXiv.org/abs/1003.0194}{\texttt{ arXiv:1003.0194}}.
Erratum \doi{10.1103/PhysRevC.82.039903}.

\bibitem{Qiu:2011iv}
\hrefCMSnoop {} {Z.~Qiu and U.~W. Heinz, ``{Event-by-event shape and flow
  fluctuations of relativistic heavy-ion collision fireballs}'',} \textit{
  Phys. Rev. C} \textbf{ 84} (2011) 024911,
  \href{http://dx.doi.org/10.1103/PhysRevC.84.024911}{\doi{10.1103/PhysRevC.84.024911}},
\href{http://www.arXiv.org/abs/1104.0650}{\texttt{ arXiv:1104.0650}}.

\bibitem{Luzum:2012wu}
\hrefCMSnoop {} {M.~Luzum and J.-Y. Ollitrault, ``{Extracting the shear
  viscosity of the quark-gluon plasma from flow in ultra-central heavy-ion
  collisions}'',} \textit{ Nucl. Phys. A} \textbf{ 904} (2013) 377,
  \href{http://dx.doi.org/10.1016/j.nuclphysa.2013.02.028}{\doi{10.1016/j.nuclphysa.2013.02.028}},
\href{http://www.arXiv.org/abs/1210.6010}{\texttt{ arXiv:1210.6010}}.

\bibitem{Gardim:2012im}
\hrefCMSnoop {} {F.~G. Gardim, F.~Grassi, M.~Luzum, and J.-Y. Ollitrault,
  ``{Breaking of factorization of two-particle correlations in
  hydrodynamics}'',} \textit{ Phys. Rev. C} \textbf{ 87} (2013) 031901,
  \href{http://dx.doi.org/10.1103/PhysRevC.87.031901}{\doi{10.1103/PhysRevC.87.031901}},
\href{http://www.arXiv.org/abs/1211.0989}{\texttt{ arXiv:1211.0989}}.

\bibitem{Heinz}
\hrefCMSnoop {} {U.~W. Heinz, Z.~Qiu, and C.~Shen, ``{Fluctuating flow angles
  and anisotropic flow measurements}'',} \textit{ Phys. Rev. C} \textbf{ 87}
  (2013) 034913,
  \href{http://dx.doi.org/10.1103/PhysRevC.87.034913}{\doi{10.1103/PhysRevC.87.034913}},
\href{http://www.arXiv.org/abs/1302.3535}{\texttt{ arXiv:1302.3535}}.

\bibitem{CMS:2008zzk}
\hrefCMSnoop {} {{ CMS} Collaboration, ``The {CMS} experiment at the {CERN}
  {LHC}'',} \textit{ JINST} \textbf{ 3} (2008) S08004,
\href{http://dx.doi.org/10.1088/1748-0221/3/08/S08004}{\doi{10.1088/1748-0221/3/08/S08004}}.

\bibitem{Grachov:2008qg}
{ CMS} Collaboration, \hrefCMSnoop {} {O.~A. Grachov {et~al.}, ``{Performance
  of the combined zero degree calorimeter for CMS}'',} in \textit{ XXX Int.
  Conf. on Calorimetry in High Energy Physics (CALOR 2008)}, M.~Livan, ed.,
  p.~012059.
\newblock 2009.
\newblock \href{http://www.arXiv.org/abs/0807.0785}{\texttt{ arXiv:0807.0785}}.
\newblock (J. Phys.: Conf. Series, 160 (2009) 012059).
\href{http://dx.doi.org/10.1088/1742-6596/160/1/012059}{\doi{10.1088/1742-6596/160/1/012059}}.

\bibitem{Lin:2004en}
Z.-W. Lin\hrefCMSnoop {} { {et~al.}, ``{A Multi-phase transport model for
  relativistic heavy ion collisions}'',} \textit{ Phys. Rev. C} \textbf{ 72}
  (2005) 064901,
  \href{http://dx.doi.org/10.1103/PhysRevC.72.064901}{\doi{10.1103/PhysRevC.72.064901}},
\href{http://www.arXiv.org/abs/nucl-th/0411110}{\texttt{
  arXiv:nucl-th/0411110}}.

\bibitem{GEANT4}
\hrefCMSnoop {} {{ Geant4} Collaboration, ``{GEANT4---a simulation toolkit}'',}
  \textit{ Nucl. Instrum. and Methods A} \textbf{ 506} (2003) 250,
\href{http://dx.doi.org/10.1016/S0168-9002(03)01368-8}{\doi{10.1016/S0168-9002(03)01368-8}}.

\bibitem{Djuvsland:2010qs}
\hrefCMSnoop {} {O.~Djuvsland and J.~Nystrand, ``{Single and Double
  Photonuclear Excitations in Pb+Pb Collisions at \rootsNN\ = 2.76 TeV at the
  CERN Large Hadron Collider}'',} \textit{ Phys. Rev. C} \textbf{ 83} (2011)
  041901,
  \href{http://dx.doi.org/10.1103/PhysRevC.83.041901}{\doi{10.1103/PhysRevC.83.041901}},
\href{http://www.arXiv.org/abs/1011.4908}{\texttt{ arXiv:1011.4908}}.

\bibitem{Lokhtin:2005px}
\hrefCMSnoop {} {I.~P. Lokhtin and A.~M. Snigirev, ``{A model of jet quenching
  in ultrarelativistic heavy ion collisions and high-\pt\ hadron spectra at
  RHIC}'',} \textit{ Eur. Phys. J. C} \textbf{ 45} (2006) 211,
  \href{http://dx.doi.org/10.1140/epjc/s2005-02426-3}{\doi{10.1140/epjc/s2005-02426-3}},
\href{http://www.arXiv.org/abs/hep-ph/0506189}{\texttt{ arXiv:hep-ph/0506189}}.

\bibitem{Khachatryan:2010gv}
\hrefCMSnoop {} {{ CMS} Collaboration, ``{Observation of Long-Range Near-Side
  Angular Correlations in Proton-Proton Collisions at the LHC}'',} \textit{
  JHEP} \textbf{ 09} (2010) 091,
  \href{http://dx.doi.org/10.1007/JHEP09(2010)091}{\doi{10.1007/JHEP09(2010)091}},
\href{http://www.arXiv.org/abs/1009.4122}{\texttt{ arXiv:1009.4122}}.

\bibitem{Chatrchyan:2011eka}
\hrefCMSnoop {} {{ CMS} Collaboration, ``{Long-range and short-range dihadron
  angular correlations in central PbPb collisions at a nucleon-nucleon center
  of mass energy of 2.76 TeV}'',} \textit{ JHEP} \textbf{ 07} (2011) 076,
  \href{http://dx.doi.org/10.1007/JHEP07(2011)076}{\doi{10.1007/JHEP07(2011)076}},
\href{http://www.arXiv.org/abs/1105.2438}{\texttt{ arXiv:1105.2438}}.

\bibitem{CMS:2012qk}
\hrefCMSnoop {} {{ CMS} Collaboration, ``{Observation of long-range near-side
  angular correlations in proton-lead collisions at the LHC}'',} \textit{ Phys.
  Lett. B} \textbf{ 718} (2013) 795,
  \href{http://dx.doi.org/10.1016/j.physletb.2012.11.025}{\doi{10.1016/j.physletb.2012.11.025}},
\href{http://www.arXiv.org/abs/1210.5482}{\texttt{ arXiv:1210.5482}}.

\bibitem{Chatrchyan:2013nka}
\hrefCMSnoop {} {{ CMS} Collaboration, ``{Multiplicity and transverse momentum
  dependence of two- and four-particle correlations in pPb and PbPb
  collisions}'',} \textit{ Phys. Lett. B} \textbf{ 724} (2013) 213,
  \href{http://dx.doi.org/10.1016/j.physletb.2013.06.028}{\doi{10.1016/j.physletb.2013.06.028}},
\href{http://www.arXiv.org/abs/1305.0609}{\texttt{ arXiv:1305.0609}}.

\bibitem{Alver:2010dn}
\hrefCMSnoop {} {B.~H. Alver, C.~Gombeaud, M.~Luzum, and J.-Y. Ollitrault,
  ``{Triangular flow in hydrodynamics and transport theory}'',} \textit{ Phys.
  Rev. C} \textbf{ 82} (2010) 034913,
  \href{http://dx.doi.org/10.1103/PhysRevC.82.034913}{\doi{10.1103/PhysRevC.82.034913}},
\href{http://www.arXiv.org/abs/1007.5469}{\texttt{ arXiv:1007.5469}}.

\bibitem{Borghini:2005kd}
\hrefCMSnoop {} {N.~Borghini and J.-Y. Ollitrault, ``{Momentum spectra,
  anisotropic flow, and ideal fluids}'',} \textit{ Phys. Lett. B} \textbf{ 642}
  (2006) 227,
  \href{http://dx.doi.org/10.1016/j.physletb.2006.09.062}{\doi{10.1016/j.physletb.2006.09.062}},
\href{http://www.arXiv.org/abs/nucl-th/0506045}{\texttt{
  arXiv:nucl-th/0506045}}.

\bibitem{glauber}
\hrefCMSnoop {} {M.~L. Miller, K.~Reygers, S.~J. Sanders, and P.~Steinberg,
  ``{Glauber modeling in high energy nuclear collisions}'',} \textit{ Ann. Rev.
  Nucl. Part. Sci.} \textbf{ 57} (2007) 205,
  \href{http://dx.doi.org/10.1146/annurev.nucl.57.090506.123020}{\doi{10.1146/annurev.nucl.57.090506.123020}},
\href{http://www.arXiv.org/abs/nucl-ex/0701025}{\texttt{
  arXiv:nucl-ex/0701025}}.

\bibitem{Alver:Glauber}
\hrefCMSnoop {} {B.~Alver, M.~Baker, C.~Loizides, and P.~Steinberg, ``{The
  PHOBOS Glauber Monte Carlo}'',} (2008).
  \href{http://www.arXiv.org/abs/0805.4411}{\texttt{ arXiv:0805.4411}}.

\bibitem{Drescher:2006pi}
\hrefCMSnoop {} {H.-J. Drescher, A.~Dumitru, A.~Hayashigaki, and Y.~Nara,
  ``{The eccentricity in heavy-ion collisions from color glass condensate
  initial conditions}'',} \textit{ Phys. Rev. C} \textbf{ 74} (2006) 044905,
  \href{http://dx.doi.org/10.1103/PhysRevC.74.044905}{\doi{10.1103/PhysRevC.74.044905}},
\href{http://www.arXiv.org/abs/nucl-th/0605012}{\texttt{
  arXiv:nucl-th/0605012}}.

\end{thebibliography}\endgroup
